\documentclass[prl,aps,epsf,twocolumn,longbibliography]{revtex4-1}
\usepackage{times}
\usepackage{graphicx}
\usepackage{float}
\usepackage{latexsym,amsmath,amssymb,bm,euscript}
\usepackage{color}
\usepackage{epstopdf}
\usepackage[colorlinks=true,linkcolor=blue,citecolor=blue,urlcolor=blue]{hyperref}

\usepackage{ulem}
\usepackage{inputenc}

\begin{document}

\title{Microscopic Theory of Superconducting Phase Diagram in Infinite-Layer Nickelates}

\author{T. Y. Xie$^{1,*}$, Z. Liu$^{2,*}$, Chao Cao$^3$, Z. F. Wang$^2$, J. L. Yang$^{2}$, W. Zhu$^4$}


\affiliation{$^1$\,\,\,\,\,\,\,\,\,\,\,\,\, Zhejiang University, Hangzhou, 310027, China\,\,\,\,\,\,\,\,\,\,\,\, \\
$^2$Hefei National Laboratory for Physical Sciences at the Microscale, University of Science and Technology of China, Hefei, Anhui 230026, China	\\
$^3$  Department of Physics, Zhejiang University,  Hangzhou, 310027, China \\
$^4$Key Laboratory for Quantum Materials of Zhejiang Province, School of Science, Westlake University, 18 Shilongshan Road, Hangzhou 310024, Zhejiang Province, China}

\begin{abstract}
Since the discovery of superconductivity in infinite-layer nickelates RNiO$_2$ (R=La, Pr, Nd), great research efforts have been paid to unveil its underlying superconducting mechanism. However, the physical origin of the intriguing hole-doped superconductivity phase diagram, characterized by a superconductivity dome sandwiched between two weak insulators, is still unclear. 
Here, we present a microscopic theory for electronic structure of nickelates from a fundamental model-based perspective.
We found that the appearance of weak insulator phase in lightly and heavily hole-doped regime is dominated by Mottness and Hundness, respectively, exhibiting a unique orbital-selective doping originated from the competition of Hund interaction and crystal field splitting. Moreover, the superconducting phase can also be created in the ``mixed" transition regime between Mott-insulator and Hund-induced insulator, exactly reproducing the experimentally observed superconducting phase diagram. Our findings not only demonstrate the orbital-dependent strong-correlation physics in Ni 3$d$ states, but also provide a unified understanding of superconducting phase diagram in hole-doped infinite-layer nickelates, which are distinct from the well-established paradigms in cuprates and iron pnictides.
\end{abstract}
\maketitle

To decipher how superconductivity (SC) emerges from normal state is a crucial step toward the physical understanding of unconventional superconductor \cite{cup0,cup1,cup2,cup3,Zaanen1985,FeSC0,Stewart2011,FeSC1}.
In early paradigms, the charge-transfer insulator \cite{Zaanen1985} and bad metal \cite{Chubukov2012} is used as parent compounds for cuprates \cite{cup1,cup3}
and iron pnictides \cite{Chubukov2012,Hund}, respectively.
Since the exotic SC mechanism is rooted in different origins of the correlation in normal state,
the exploration of new paradigm for SC is of great importance, which could further enrich the zoology of
unconventional SC in strongly-correlated materials.
The discovery of SC in infinite-layer nickelates RNiO$_2$ (R=La, Pr, Nd) \cite{Nd0, Nd1, Nd2, Nd3, La01, La02, Pr0, Pr1, Pr2}
offers a new platform for investigating the mechanism of unconventional SC.
Especially, there are two key features in its experimental SC phase diagram, which are absence in cuprates and iron pnictides:
i)  weak insulator in both lightly and heavily hole-doped regimes \cite{Nd1, Nd2, La01, La02, Pr1};
ii) SC dome sandwiched between two weak insulator regimes \cite{Nd1, Nd2, La01}.
Currently, the origin of this anomalous SC phase diagram remains outstanding. 
It is highly desirable to explore the strong-correlation physics behind this SC phase diagram,
and make a possible connection to or distinction from the well-established SC mechanisms in cuprates and iron pnictides.

\begin{figure*}
	\includegraphics[width=0.85\textwidth]{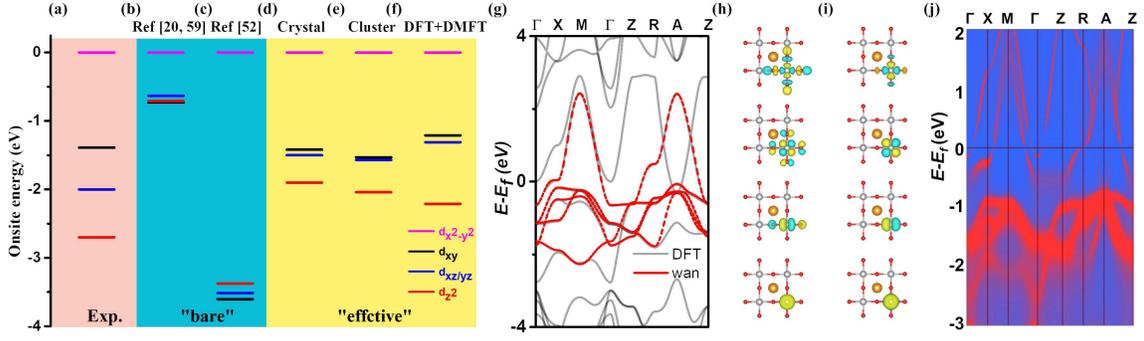}
	\caption{Crystal field splitting (CFS) for Ni 3$d$ orbitals in infinite-layer nickelates. (a) Data from RIXS experiment \cite{CFS0}. The ``bare" 3d orbital sequence from (b) Ref. \cite{bandNd0,CFS2} and (c) Ref. \cite{cal0}. Onsite energies of ``effective" 3d orbitals from (d) the crystal method (see text), (e)  the cluster method, (f) DMFT calculation. (g) The DFT band structure and Wannier fitted effective bands within the crystal method. Maximally localized Wannier functions for (h) ``effective" Ni 3$d$ orbitals, compared with (i) ``bare" Ni 3$d$ and O 2$p$ orbitals. (j) Momentum-resolved spectral function from DFT+DMFT calculations for LaNiO$_2$ at 116 K.
	} \label{fig:CFS}
\end{figure*}

Although the SC mechanism in infinite-layer nickelates is a controversial topic, there has been several theoretical consensus for its electronic structures.
The normal state is more proximate to a Mott-Hubbard insulator \cite{bandNd0, bandNd1, bandNd2, bandNd3, Jiang2020}.
The correlation in Ni 3$d_{x^2-y^2}$ orbital is relevant to SC because of the structure analogy to cuprates \cite{1band1, 1band2}.
The Ni 3$d$ states are influenced by a self-doping rare-earth-orbital band (served as a charge reservoir) through hybridization effect \cite{kondo1, kondo2,Lu2021}.
However, the appearance of itinerant electronic band cannot interpret the weakly insulating phase in heavily hole-doped regime \cite{Adhikary2020, Nica2020, Peng2021}.
Very recently, intensive studies have also demonstrated the importance of multi-Ni-orbital nature and concomitant Hund's interaction \cite{multi0, multi1, multi2, multi3, multi4, multi5, multi6, multi7, multi8, multi9, multi10,multi11,multi12}.
Nevertheless, the role of multi-orbitals in SC phase diagram is still under debate.
Taken as a whole, despite of various works on normal state properties, a complete and unified physical understanding of the experimental SC phase diagram upon hole-doping remains unexplored.

In this work, driven by recent x-ray experimental observations \cite{CFS0} and first-principles calculations,
we build a microscopic two-band Hubbard model with Ni $\{3d_{x^2-y^2}$, 3$d_{xy}\}$ orbitals.
Based on mean-field calculations and interplay analysis of Hund interaction ($J_H$)
and crystal field splitting ($\eta$), we directly identify a theoretical SC phase diagram with remarkable features:
    i) weak insulator phase dominated by orbital-selective Mottness-like physics in lightly hole-doped regime,
	ii) weak insulator phase dominated by moderate $J_H$ selected Hundness-like physics in heavily hole-doped regime,
	iii) SC phase dominated by $d$-wave paring between two weak insulators in an optimal hole-doped regime.
Our results provide a microscopic model and unified physical picture for describing the electronic structures
and understanding the SC phase diagram in nickelates \cite{Nd0, Nd1, Nd2, Nd3, La01, La02, Pr0, Pr1, Pr2},
that is, being a moderately correlated system, the combined effect of orbital-selective Mottness and Hundness
makes nickelate-family a bridge connecting cuprates and iron pnictides.

\textit{First-principals analysis.---}
To construct a reliable microscopic model of nickelates,
an accurate description of its crystal field splitting (CFS) is the first step,
which will shed lights on bonding nature and put strong constrains on model.
Recently, the experimental measurement of CFS in nickelates has been exploited by resonant inelastic x-ray scattering (RIXS) \cite{CFS0}, reporting an orbital-sequence of $d_{x^2-y^2}$ (0 eV) $>d_{xy}$ ($-$1.39 eV) $>d_{xz}/d_{yz}$ ($-$2.0 eV) $>d_{z^2}$ ($-$2.7 eV) (see Fig. \ref{fig:CFS}(a)).
However, this significant observation cannot be simply explained by the ``bare" Ni $3d$ orbitals in previous first-principles calculations. For example, the result from Botana et al. \cite{bandNd0}, Hepting et al. \cite{CFS2} is shown in Fig. \ref{fig:CFS}(b) and that from Jiang et al. \cite{cal0} is shown in Fig. \ref{fig:CFS}(c), both of which significantly deviate from that shown in Fig. 1(a).

Due to such a disagreement between theory and experiment, we use three different methods to check the CFS in detail. First, the CFS is calculated through Wannier downfolding, dubbed as "crystal" method.
Using five effective Ni 3$d$ orbitals to fit the first-principles band structures (Fig. \ref{fig:CFS}(g)),
crucially, the obtained orbital-sequence is consistent with experiment (Fig. \ref{fig:CFS}(d)).
The spatial distribution of these Wannier functions (WFs) has contributions from both
``bare" Ni $3d$ orbitals and O $2p$ orbital (Fig. \ref{fig:CFS}(h)). Since the WFs carry more information from
high-energy orbitals \cite{Goodge2021, Shen2021}, the associated Wannier Hamiltonian is similar to an effective low-energy one.
As a comparison, if more orbitals are included in the fitting process  see Fig. \ref{fig:CFS-Ni-1} and Fig. \ref{fig:CFS-Ni-2}), the WFs are closer to atomic orbitals (Fig. \ref{fig:CFS}(i)),
making the associated Wannier Hamiltonian similar to a "bare" one.
Second, the CFS is calculated through cluster model proposed by Eskes \textit{et al.} \cite{Eskes1990},
please see details in Supplementary Materials Sec. A-3 \cite{sm}.
The obtained orbital-sequence is also consistent with experiment (Fig. \ref{fig:CFS}(e)).
Importantly, this method allows us to quantitative analyze the components of effective orbitals.
Taking $3d_{x^2-y^2}$ as an example, the weight of O $2p$ orbital in this effective orbital of NdNiO$_2$ is $\sim23.8\%$,
which is nearly half of that in CaCuO$_2$ ($\sim44.8\%$) \cite{sm}. 
This analysis, complementary with the "crystal" method, well explains the components of effective Ni $3d$ orbitals observed in experiment. Third, the CFS is calculated through DFT+DMFT, which is comparable to recent many-body quantum chemistry method \cite{CASSCF1}. Fig. \ref{fig:CFS}(j) shows the momentum-resolved spectral function of DFT+DMFT, where the extracted orbital-sequence is  $d_{x^2-y^2}$ (0 eV) $>$ $d_{xy}$ ($-$1.21 eV) $>$ $d_{zx/zy}$ ($-$1.31 eV) $>$ $d_z^2$ ($-$2.21 eV) (Fig. \ref{fig:CFS}(f)).
This result not only has qualitatively the same sequence as, but also is numerically close to the experiment (Fig. \ref{fig:CFS}(a)).
Physically, the above CFS can be understood in a simple picture.
Due to $D_{4h}$ symmetry of nickelates, the out-of-plane orbitals $\{d_{z^2} , d_{xz}, d_{yz}\}$ have lower energies by extending orbital along c-axis, leaving in-plane orbitals \{$d_{x^2-y^2}$, $d_{xy}$\} more relevant to Fermi level \cite{multi0}, akin to the case in the infinite-layer cuprate CaCuO$_2$ (see Supple. Mat.  for more details\cite{sm}).

\begin{table}[t]
	\caption{Two-band model parameter for $\rm RNiO_2$ (R=La, Pr, Nd). $\epsilon(1)$ and  $\epsilon(2)$ are onsite energy for $d_{x^2-y^2}$ and $d_{xy}$ WFs. $t(1)$ \& $t(2)$ ($t'(1)$ \& $t'(2)$) are in-plane nearest (next-nearest) neighbor hopping strength for $d_{x^2-y^2}$ and $d_{xy}$ WFs. $U$ ($U'$) and $J_H$ represent the intra-(inter-)orbital Coulomb repulsion and Hund's coupling. }
	\label{tab:occf}
	\begin{tabular}{c|c|c|c|c|c|c|c|c}
		\hline\hline
		R  & $\epsilon(1)-\epsilon(2)$ & $t(1)$ & $t'(1)$ & $t(2)$ & $t'(2)$ & $U$ & $U'$ & $J_H$ \\
		\hline
		La & 1.39 & -0.37 & 0.10 & -0.16 & -0.05 & 3.60 & 1.90 & 0.84 \\
		Pr & 1.41 & -0.37 & 0.09 & -0.16 & -0.05 & 3.63 & 1.94 & 0.84\\
		Nd & 1.42 & -0.37 & 0.09 & -0.16 & -0.05 & 3.64 & 1.95 & 0.84\\
		\hline\hline
	\end{tabular}
\end{table}

The above three different methods give the same CFS with the experimental observations \cite{CFS0},
indicating the in-plane Ni \{$3d_{x^2-y^2}$, $3d_{xy}$\} orbitals to be more relevant to the Fermi level.
With these considerations, we propose a two-band microscopic model as:
\begin{equation*}
\begin{split}
	\hat{H}_{TB} &= \sum_{i,\alpha,\sigma} \epsilon(\alpha) \hat{d}^{\dagger}_{i\alpha\sigma}\hat{d}_{i\alpha\sigma} + \sum_{\langle i,j \rangle,\alpha\sigma}(t(\alpha)\hat{d}^{\dagger}_{i\alpha\sigma}\hat{d}_{j\alpha\sigma} + h.c.) \\
	& + \sum_{\langle\langle i,j \rangle\rangle,\alpha\sigma}t'(\alpha)\hat{d}^{\dagger}_{i\alpha\sigma}\hat{d}_{j\alpha\sigma} + h.c.\\
\end{split}
\end{equation*}
where $\sigma$ is spin index, $i$ and $\alpha$ is site- and orbital-index for 3$d_{x^2-y^2}$ and 3$d_{xy}$ WFs. $\langle...\rangle$ and $\langle\langle...\rangle\rangle$ represent the nearest and next-nearest neighbor (NN $\&$ NNN) hopping. Since the 3$d_{x^2-y^2}$ and 3$d_{xy}$ are almost orthogonal to the rest 3$d$ WFs, these model parameters are directly extracted from the crystal model (see \cite{sm}). The out-of-plane hopping value of $d_{x^2-y^2}$ ($d_{xy}$) WF is only 10\% (20\%) of its in-plane value,
so the system shows a quasi-two-dimensional (2D) nature. Therefore, we consider only the hopping within the effective
quasi-2D NiO$_2$ plane. As shown in Tab. \ref{tab:occf}, both the   NN and NNN hopping parameter of $d_{x^2-y^2}$ is twice larger than that of $d_{xy}$, giving an opportunity to see the orbital-selective physics as we show below.

In order to investigate the interactions in nickelates, we consider the following Hamiltonian \cite{Hund}:
\begin{equation*}
\begin{split}
\hat{H}_{int} = U \sum_{i, \alpha} \hat{n}_{i\alpha\uparrow} \hat{n}_{i\alpha\downarrow}
+ \sum_{i, \sigma, \sigma'} (U'-J_H\delta_{\sigma\sigma'})\hat{n}_{i1\sigma}\hat{n}_{i2\sigma'}
\end{split}
\end{equation*}
where $U$ ($U'$) denotes intra-(inter-)orbital Coulomb repulsion, and $J_H$ denotes Hund's coupling. We take $U'=U-2J_H$ so that the Hamiltonian is rotationally invariant in the orbital space. The $U$ ($U'$) and $J_H$ on 3$d$ WFs are estimated from the
first-principles calculations with constrained random phase approximation (cRPA) \cite{cRPA1, cRPA2, cRPA3}. The interaction parameters for three nickelates are listed in Tab. \ref{tab:occf}, showing the similar strength with well-kept relationship $U'=U-2J_H$. The small value of $U$ and relatively large $J_H$ therefore puts the infinite-layer nickelates as a moderate correlated system closer to iron-pnictides \cite{UFe} than cuprates.

\begin{figure}[t]
	\includegraphics[width=0.9\linewidth]{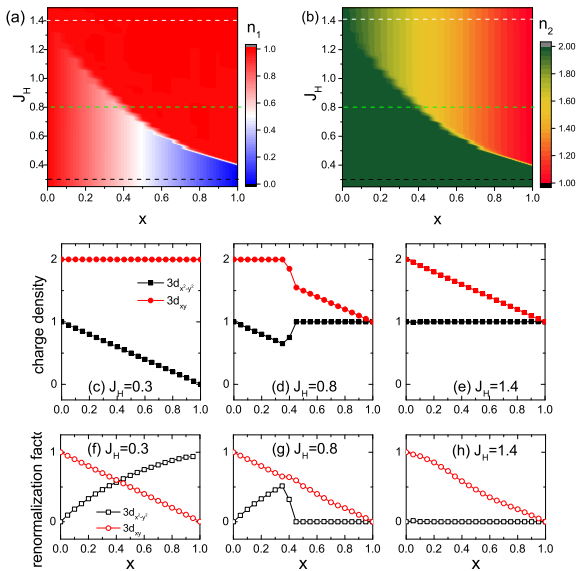}
	\caption{Evolution of many-body electronic structure with hole-doping.
		Density plot of orbital occupation on (a) Ni 3$d_{x^2-y^2}$ orbital and (b) Ni 3$d_{xy}$ orbital in the two-band Hubbard model, as a function of Hund's interaction $J_H$ and doping ratio $x$. 
		(c-e) Orbital-resolved charge density evolution as a function of  $x$. The three different panels respectively corresponds to different line cuts in subfig (a,b): (c) $J_H=0.3$ eV, (d) $J_H=0.8$ eV, (e) $J_H=1.4$ eV.
		(f-h) Orbital-resolved renormalization factor $Z_\alpha$ (quasi-particle weight) as a function of $x$, for (f) $J_H=0.3$ eV, (g) $J_H=0.8$ eV, (h) $J_H=1.4$ eV.
		Here we set the parameters $t_1=0.375$ eV, $t_2=0.15$ eV, $U_0=3.5$ eV, $\eta=1.2$ eV.
	}
	\label{fig:sb}
\end{figure}

\textit{Two weakly insulators.---}
Having established the microscopic two-band Hubbard model with orbitals relevant to low-energy physics of nickelates,
we first consider the evolution of its electronic structures upon hole-doping.
We introduce a slave-boson formalism \cite{Kotliar1986,Kotliar1988} to decouple the exchange interactions,
using a direct multi-orbital generalization of original single-orbital scheme \cite{Sigrist2005} (details see supple. mat. \cite{sm}).
Fig. \ref{fig:sb}(a-b) presents the orbital-resolved charge density as a function of Hund's interaction ($J_H$) and doping ratio ($x$).
We focus on $0\leq x\leq 1$ that corresponds to the hole-doping evolution from $3d^9$ to $3d^8$ configuration on NiO$_2$ plane.
The main feature is that there are three distinct phases depending on the strength of $J_H$ versus $\eta$, called as orbital-selective Mottness regime, Hundness regime and ``mixed" regime. The Mottness (Hundness) phase occupies the small (large) $J_H$ regime, while the mixed phase emerges in between.
Fig. \ref{fig:sb}(c-e) present the doping dependent orbital-resolved charge density ($n_\alpha$) in three different regimes.
In the orbital-selective Mottness regime  (Fig. \ref{fig:sb}(c)),
the doped-holes reside on Ni 3$d_{x^2-y^2}$ orbital, and the Ni 3$d_{xy}$ orbital is totally-filled.
Since a strong crystal field ($\eta$) favors a large orbital polarization,
holes tend to fill the 3$d_{x^2-y^2}$ orbital in a low-spin configuration.
In the Hundness regime (see Fig. \ref{fig:sb}(e)),
the doped-holes reside on the Ni 3$d_{xy}$ orbital only.
This is the result of a large Hund's exchange ($J_H$) promoting the
carriers on different orbitals in a high-spin state to minimize repulsive interactions.
Importantly, in the mixed phase (Fig. \ref{fig:sb}(d)), doping leads to a transition from Mottness to Hundness, 
where the holes reside on 3$d_{x^2-y^2}$ orbital in the regime $x<x^*$, while the holes begin to populate 3$d_{xy}$ orbital in the regime $x>x^*$.
Here, the critical value of $x^*$ depends on $J_H,\eta$, i.e.
the larger (smaller) $J_H$ ($\eta$), the smaller value of $x^*$.

Furthermore, the competition between Hundness and Mottness can be revealed by the renormalization factor $Z_\alpha$ (i.e. inverse of effective mass $\sim m^{-1}_{\alpha}$) of two bands, 
which qualifies the effective carrier quasi-particle weight, as shown in Fig. \ref{fig:sb}(f).
In the orbital-selective Mottness regime, Ni 3$d_{x^2-y^2}$ orbital is active and its quasiparticle weight increases as the hole-doping.
In the Hundness regime (Fig. \ref{fig:sb}(h)), 3$d_{x^2-y^2}$ orbital is locked by Hund's interaction thus quasiparticle weight is pinned at exactly zero.
In the mixed regime (Fig. \ref{fig:sb}(g)),  $Z_{x^2-y^2}$ exhibits a non-monotonic behavior, with a maximum around $x\sim x^*$.
After the Hundness physics sets in ($x<x^*$), $Z_{x^2-y^2}$ drops to zero.

\begin{figure}[t]
	\includegraphics[width=0.58\linewidth]{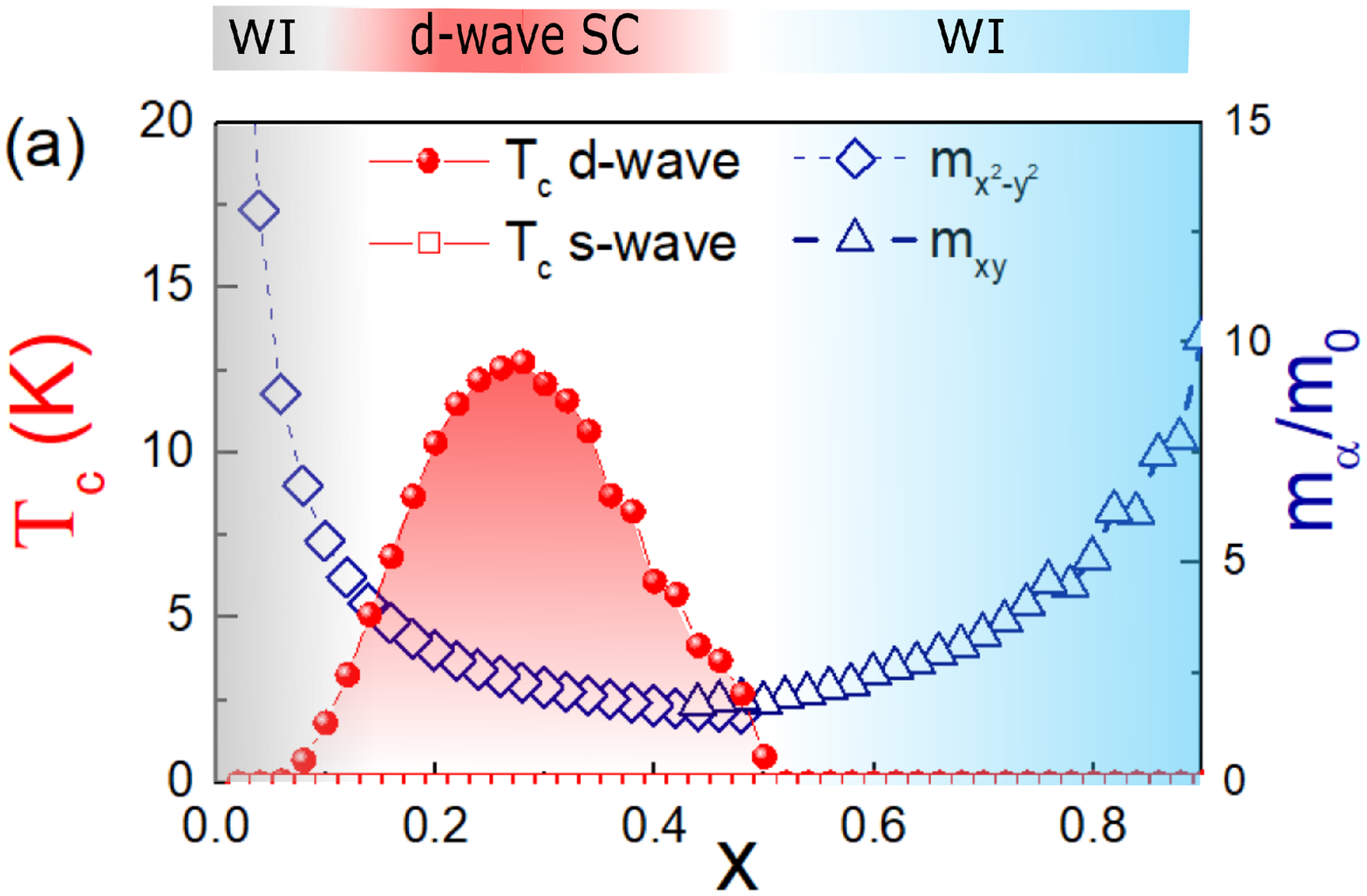}
	\includegraphics[width=0.41\linewidth]{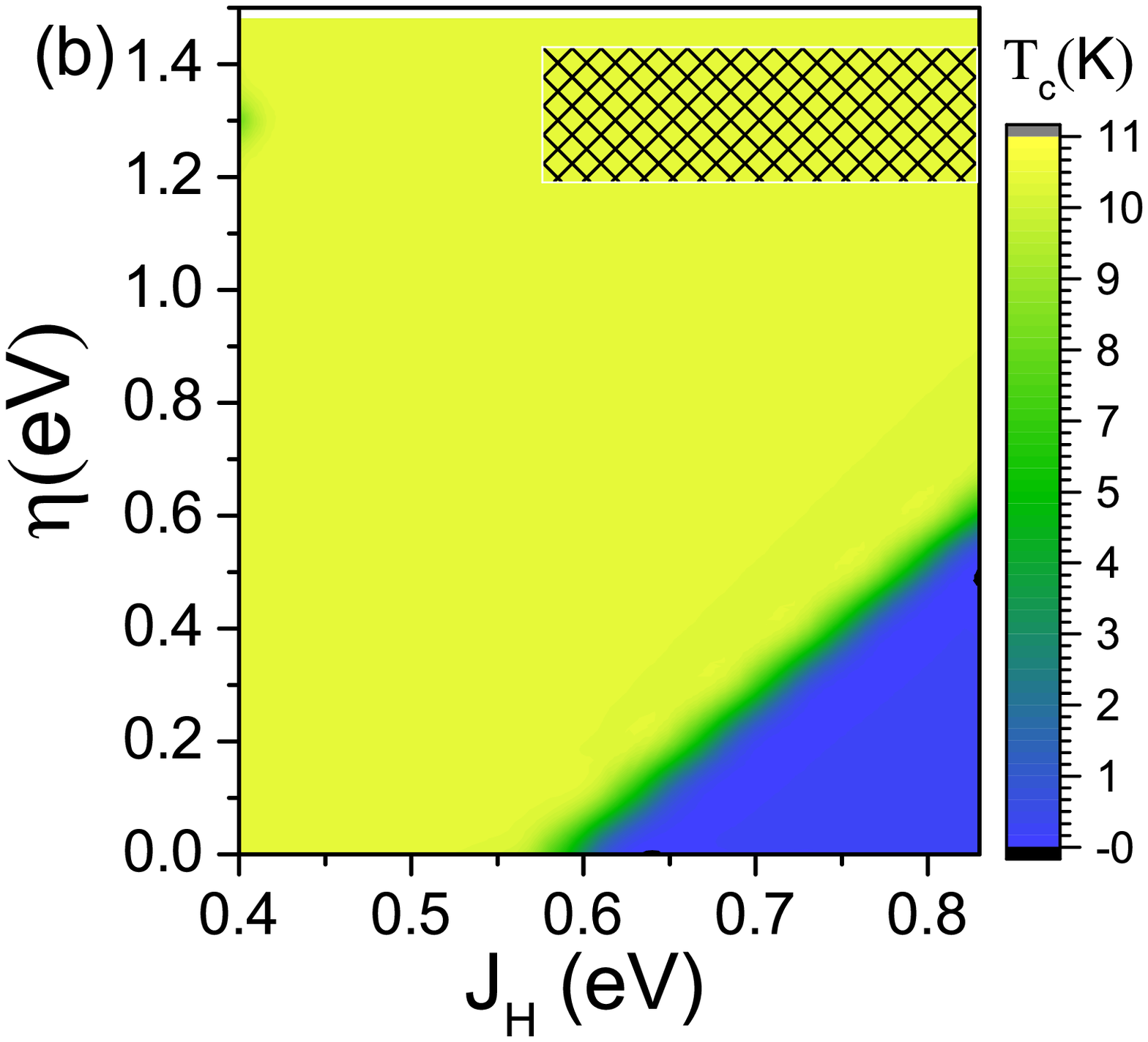}
	\caption{ Phase diagram upon hole-doping, including d-wave SC and two weakly insulators (WIs).
		(a) $T_c$ versus doping ratio $x$ for s-wave (red square) and d-wave (red dots) pairing symmetry,
		and effective mass $m_\alpha/m_0$ for 3d$_{x^2-y^2}$ (blue diamond) and 3d$_{xy}$ (blue triangular) orbital upon doping ratio $x$.
		Here we set $J= 0.20$ eV, $t_1=0.375$ eV, $t_2=0.15$ eV, $U_0=3.5$ eV, $J_H=0.7$ eV, $\eta=1.2$ eV.
		(b) Heatmap of $T_c$ versus $J_H$ and $\eta$, by setting $x=0.2$,  $J= 0.20$ eV, $U_0=3.5$ eV.
		The shaded region marks parameters relevant to nickelates.
	}
	\label{fig:sc}
\end{figure}

Here we stress that the ``mixed" phase exhibits weakly insulating behavior in both lightly and heavily hole-doped regimes (Fig.\ref{fig:sc}(a)), but the origin of them is different.
In the lightly hole-doped regime, the insulator comes from the suppressed kinetic mobility of carriers on 3$d_{x^2-y^2}$ WF and vanishing small carrier density on 3$d_{xy}$ WF.
While in the heavily hole-doped regime, the insulating behavior is produced by frozen carriers on 3$d_{x^2-y^2}$ and strong correlation due to Hundness.
Thus, we conclude that the ``mixed" phase induced by the competition between $J_H$ and $\eta$ leads to insulating behavior in both lightly and heavily hole-doped regimes, providing a natural understanding of experimental SC phase diagram.

\textit{Superconductivity.---}
Last we turn to study the SC in our model.
We assume the carrier pairing is mediated by the spin fluctuations,
and additional anti-ferromagnetic interactions between the moment of charge carriers survive \cite{Lu2021, Zhou2021, Lin2021, Hepting2021}
$H_{int} = J \sum_{\langle ij \rangle,\alpha} \mathbf{S}_{i} \cdot \mathbf{S}_{j} - \frac{1}{4}n_{i} n_{j}$,
where $J$ denotes the effective spin exchange strength between 3$d_{x^2-y^2}$ orbitals.
Then we treat this interaction at the mean-field level, and self-consistently solve the pairing strength and critical temperature in the Bogoliubov-de Gennes (BdG) equations (see  \cite{sm}).
Fig. \ref{fig:sc}(a) shows the critical temperature $T_c$ as a function of hole ratio.
We find  $T_c$ of extended s-wave symmetry is vanishing small, and nonzero $T_c$ is present for d-wave symmetry in the underdoped regime.
That a SC dome appears  sandwiched between orbital-selective Mott-insulator and Hund-induced insulator regime,
agrees with the experimental observations.

To see the robustness of SC, other $\eta$ and $J_H$ values are studied here. Due to the screening effect from other 3d WFs, the realistic $J_H$ between 3$d_{x^2-y^2}$ and 3$d_{xy}$ may deviate from the value listed in Tab. \ref{tab:occf} up to $30\%$. And the three different models in determining CFS allows a reasonable window for $\eta$. 
In Fig. \ref{fig:sc}(b), it is clear the SC is stable in the parameter region relevant to nickelates (Fig. \ref{fig:sc}(b)). In this regard we conclude the SC is robust and insensitive to the values of $J_H,\eta$.

\textit{Conclusion.---}
Using comprehensive many-body computations based on a first-principles microscopic Hamiltonian,
we present a unified physical picture for understanding the hole-doping superconducting phase diagram in
infinite-layer nickelates \cite{Nd0, Nd1, Nd2, La01, La02, Pr0, Pr1}, and
provide a quantitative basis for theoretical models in describing the electronic structure revealed in RIXS \cite{CFS0}.
Our study implies that infinite-layer nickelate-based superconductors, in lightly hole-doped regime,
are analog to the cuprates with active 3$d_{x^2-y^2}$ orbital, resulting in Mottness physics.
In contrast, in the heavily hole-doped regime, it shares many similarities with
iron-based superconductors, such as the importance of Hund's interaction and tendency toward high-spin configurations.
In this context, infinity-layer nickelate is a moderately correlated system
in which the electronic structures of the NiO$_2$ layer bears similarities to those in either cuprates or iron-based materials in different regions.
To further support the above picture, the hole-doped Nd$_6$Ni$_5$O$_{8}$ compound can be studied (un-doped Nd$_6$Ni$_5$O$_{8}$ is equivalent 
to 3d$^{8.8}$ configuration \cite{Pan2021,Botana2021}), and a weak insulator phase is expected in its heavily hole-doped regime.
In addition, important future problems also include the exploration of possible enhancement of superconductivity in such a multi-orbital system.

\textit{Acknowledgments.---}W.Z. thanks H. H. Chen, J. H. Dai, K. Jiang, M. Jiang, Q. Y. Lu, F. Lechermann, C. A. Lane, Q. H. Wang, X. G. Wan, C. J. Wu, J. Wu,  Y. F. Yang, G. M. Zhang, and J. X. Zhu for discussion.
W.Z. thanks M. R. Norman for critical comments.
This work was supported by ``Pioneer" and ``Leading Goose" R\&D Program of Zhejiang (2022SDXHDX0005), the Key R\&D Program of Zhejiang Province (2021C01002) and the foundation from Westlake University. 
Z.F.W. was supported by NSFC (No. 12174369, 11774325), National Key Research and Development Program of China (No. 2017YFA0204904) and Fundamental Research Funds for the Central Universities.


$^*$ These two authors contributed equally.

\clearpage
\appendix
\widetext
\begin{center}
	\textbf{\large Supplementary Materials for: \\`` Microscopic Theory of Superconducting Phase Diagram in Infinite-Layer Nickelates ''}
\end{center}

\vspace{1mm}

\renewcommand\thefigure{\thesection S\arabic{figure}}
\renewcommand\theequation{\thesection S\arabic{equation}}

\setcounter{figure}{0}
\setcounter{equation}{0}

In this supplemental materials, we provide some more numerical results to support the conclusions we have discussed in the main text. In Sec. A, we present the computational details of density-functional theory and dynamical mean-field theory calculations, and make a comparison with  CaCuO$_2$. In Sec. B, we present an introduction of the slave-boson method for mean-field calculations. In Sec. C, we outline the Bogoliubov-de Gennes equation for superconductivity used in this work. In Sec. D, we provide further discussion to understand the contribution of R 5$d$ electrons from rare-earth element.

\section*{A. DFT simulations}

\subsection*{A-1. Crystal structure}
\begin{figure*}[b]
	\includegraphics[width=0.8\textwidth]{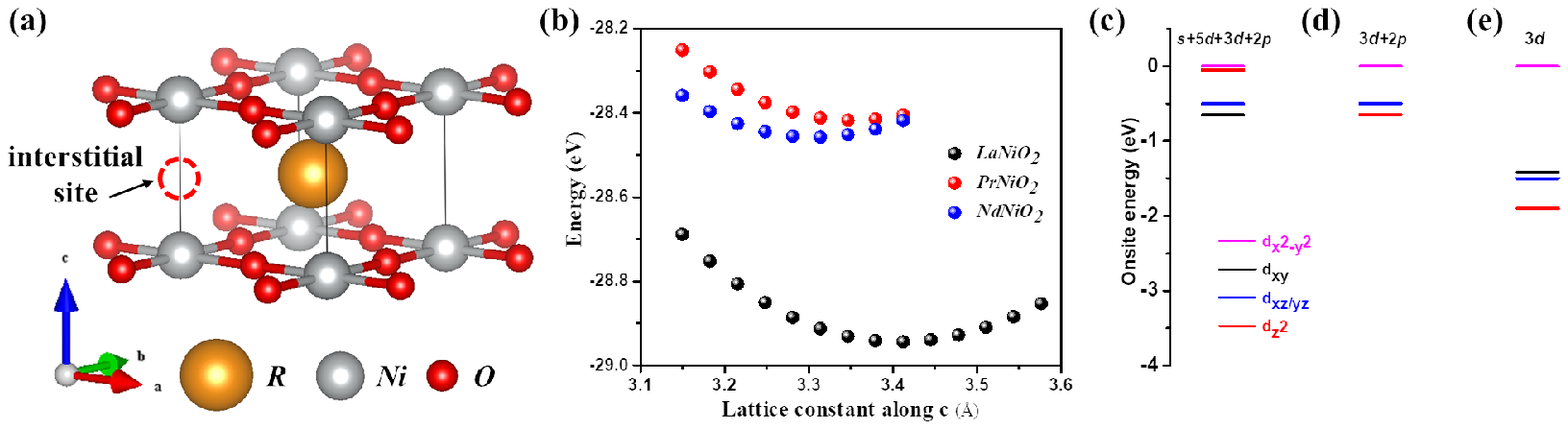}
	\caption{(a) Archetype structure of RNiO$_2$, R = La, Pr, Nd. The interstitial site is marked by the red dashed circle. (b) Energy versus lattice constant in c direction for RNiO$_2$. (c) Onsite energies of Ni 3$d$ WFs when interstitial $s$, Nd 5$d$, Ni 3$d$ and O 2$p$ orbitals are chosen in the downfolding. (d) Onsite energies of Ni 3$d$ WFs when Ni 3$d$ and O 2$p$ orbitals are chosen in the downfolding. (e) Onsite energies of Ni 3$d$ WFs when only Ni 3$d$ orbitals are used in the downfolding. Here (e) is plotted for a clearer comparison.
	\label{fig:structure}}
\end{figure*}

Density functional theory (DFT) calculations are performed within the plane wave, projector augmented wave method as implemented in the Vienna \textit{ab initio} simulation package VASP \cite{vasp1, vasp2, vasp3}. The generalized gradient approximation was used for the exchange-correlation potential \cite{vasp4}. The infinite layered structure ABO$_2$ can be regarded as obtaining from cubic perovskite ABO$_3$ by removing apical oxygen atoms (the left vacancy site is called interstitial site as shown in Fig. \ref{fig:structure}(a)). To simulate the growth of $\rm RNiO_2$ (R=La, Pr, Nd) layers on substrate $\rm SrTiO_3$, the in-plane lattice constant of $\rm RNiO_2$ is fixed to that of $\rm SrTiO_3$ at 3.92 {\AA}. The out-of-plane parameter is scanned to obtain the optimal value (the potential energy surface is shown in Fig. \ref{fig:structure}(b)), which is 3.41, 3.35 and 3.31 {\AA} for LaNiO$_2$, PrNiO$_2$ and NdNiO$_2$. Because of the removing of apical O, the lattice constant in the c direction is much smaller than the in-plane lattice constants.

\subsection*{A-2. Wannier downfolding}
To obtain parameters such as onsite energy and hopping integral, we downfold the full Hamiltonian into the subspace in Wannier90 package\cite{Wanpac}. The downfolding process also allows us to obtain the following matrix element:
\begin{equation}
H_{\alpha\beta}(R) = <\phi_{0,\alpha}|\hat{H}|\phi_{R,\beta}>
\end{equation}
where $|\phi_{0,\alpha}>$ is the maximally localized Wannier function $\alpha$ in home cell (index as 0) and  $|\phi_{R,\beta}>$ the maximally localized Wannier function $\beta$ in cell R. When R=0, $\alpha=\beta$, the above matrix element orbital energy, otherwise we obtain the hopping integral.

\begin{figure*}
	\includegraphics[width=0.8\textwidth]{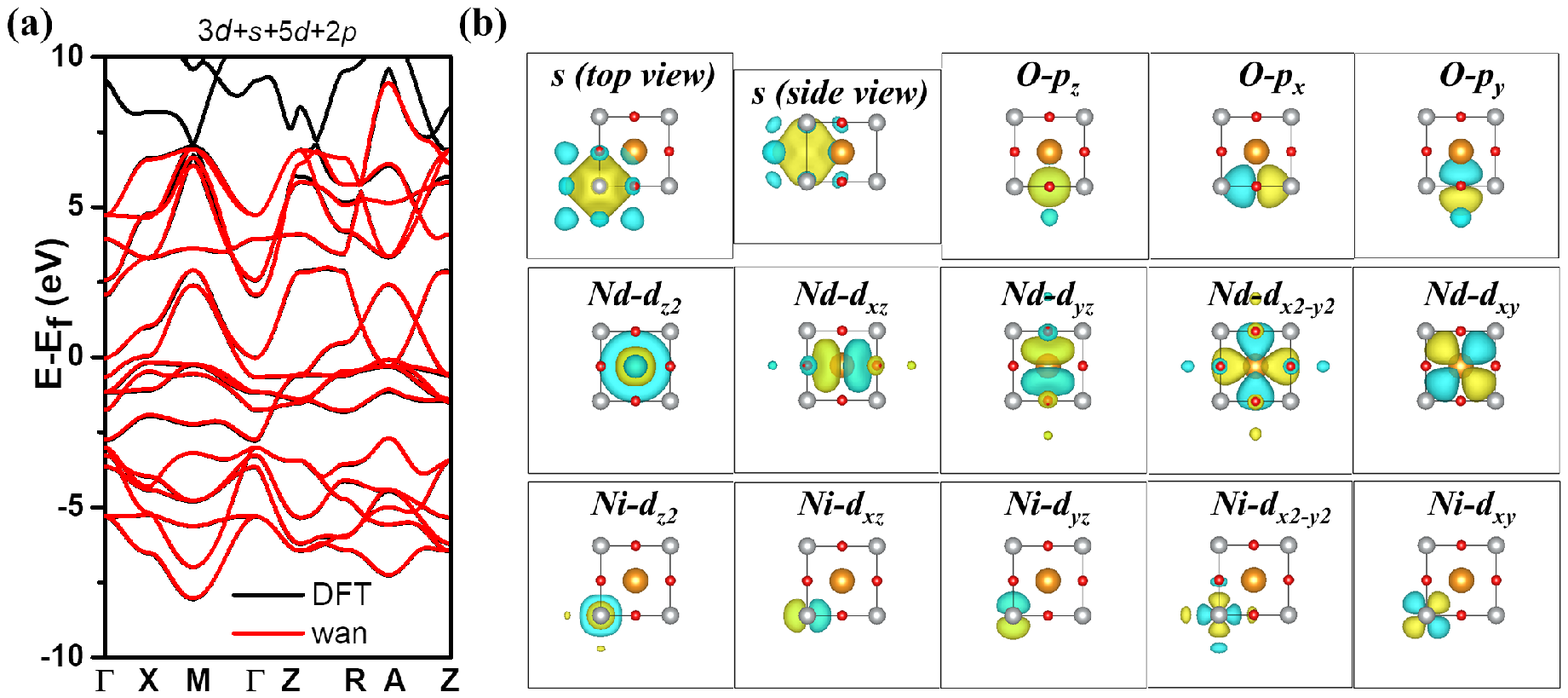}
	\caption{Downfolding in the "bare" limit. (a) The DFT band structure and the Wannier fitted bands. (b) Maximally localized Wannier functions. The WFs on the other oxygen atoms is ignored here because of symmetry.
	\label{fig:CFS-Ni-1}}
\end{figure*}

For example, the subspace can be chosen as interstitial $s$, Nd 5$d$, Ni 3$d$ and O 2$p$ orbitals. There are 17 orbitals in total. The Wannier fitted band structure with respect to first-principles calculations is shown in Fig. \ref{fig:CFS-Ni-1}(a) and the obtained WFs are displayed in Fig. \ref{fig:CFS-Ni-1}(b). From Fig. \ref{fig:CFS-Ni-1}(a), the fitted band structure is exactly the same as DFT in a very large energy window and WFs in Fig. \ref{fig:CFS-Ni-1}(b) are very close to the corresponding atomic orbitals, so it is reasonable to call the Hamiltonian obtained here as "bare" one (The real bare Hamiltonian should contain other bands including core levels, Ni 3$s$, 3$p$ and empty ones. These are quite high in energy and only renormalize the parameters by a small amount. Therefore it is safe to ignore these bands and call the Hamiltonian of 17 bands as bare Hamiltonian). In this limit, the obtained onsite energy of Ni 3$d$ WFs is shown in Fig. \ref{fig:structure}(c). As $\{3d_{x^2-y^2}, 3d_{z^2}\}$ and $\{3d_{xy}, 3d_{xz}, 3d_{yz} \}$ are almost degenerate,  the Ni atoms now have coordination environments close to $O_h$ spatial group.

\begin{figure*}
	\includegraphics[width=0.8\textwidth]{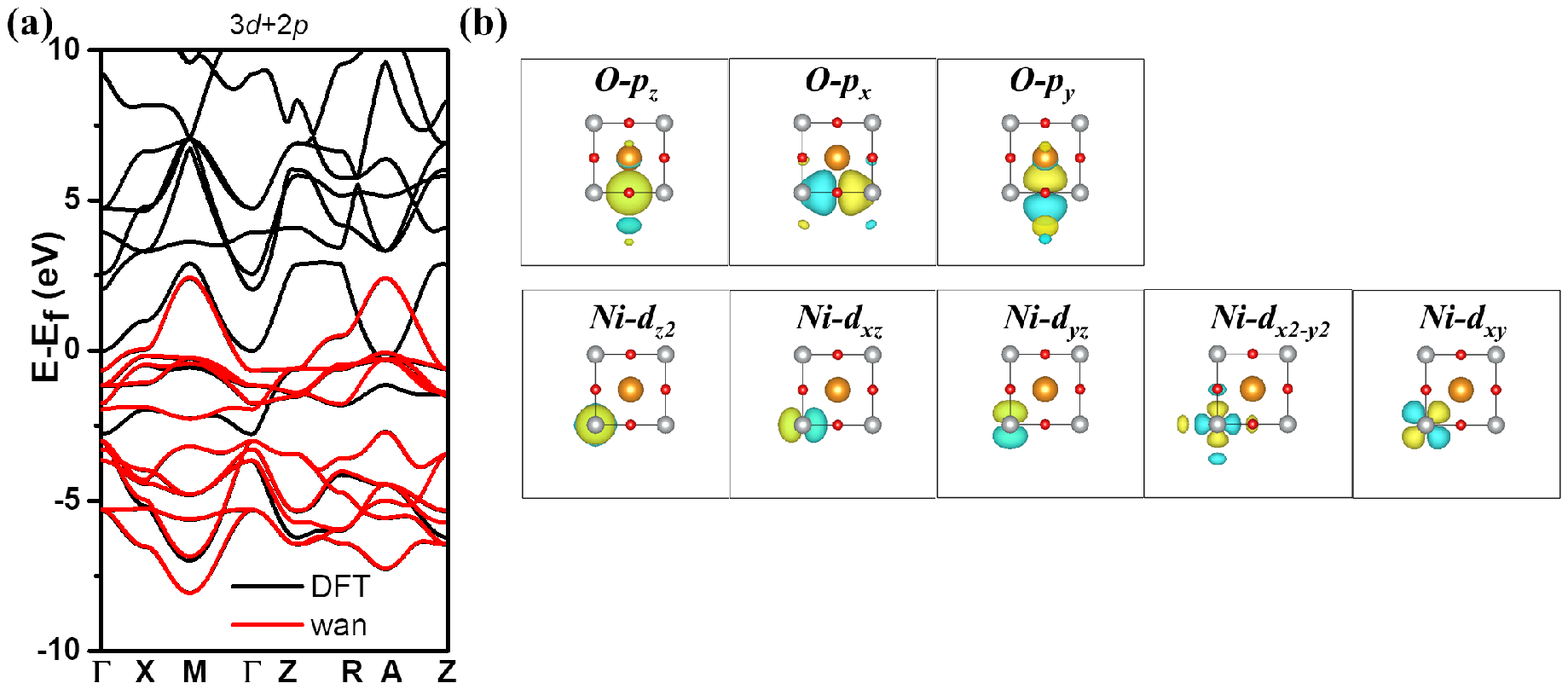}
	\caption{Downfolding with Ni 3$d$ and O 2$p$ orbitals. (a) The DFT band structure and the Wannier fitted bands. (b) Maximally localized Wannier functions. The WFs on the other oxygen atoms is ignored here because of symmetry.
	\label{fig:CFS-Ni-2}}
\end{figure*}

We can reduce the number of bands in the downfolding, then the contributions of these abandoned bands are projected to the kept subspace. Here we abandon higher energy bands: interstitial $s$ orbital and Nd 5$d$, so the effective Hamiltonian now contains 11 bands: five Ni 3$d$ and six O 2$p$ WFs. The Wannier fitted band structure with respect to first-principles calculations is shown in Fig. \ref{fig:CFS-Ni-2}(a) and the obtained WFs are displayed in Fig. \ref{fig:CFS-Ni-2}(b). Since there is large interaction between Ni 3$d_{z^2}$, interstitial s and Nd 5$d_{z^2}$, the abandon of interstitial s and Nd 5$d_{z^2}$ in the downfolding will be reflected on Ni 3$d_{z^2}$ WF. As shown in Fig. \ref{fig:structure}(d), although the onsite energy of the other four 3$d$ WFs does not change, the onsite energy of 3$d_{z^2}$ is largely reduced and close to 3$d_{xy}$.

Furthermore, in the downfolding process, we can construct ``effective" (five) Ni 3$d$ orbitals only, dubbed as the crystal model (compared with cluster method as shown below).
In practice, this is equivalent to choosing subspace as (five) Ni 3$d$ orbitals only in the Wannier downfolding. And the obtained on-site energy for Ni 3$d$ orbitals is shown in Fig. 1(d) of the main text. One sees that the onsite energy of 3$d_{z^2}$ is further reduced. Please note that  the obtained orbitals contain contributions from both "bare" 3d orbitals and 2p orbitals, so that we call them "effective" orbitals to distinguish them from the "bare" ones.

\begin{table}[t]
	\caption{Cluster model parameter. Here $d_{yz}$ and $\frac{1}{\sqrt{2}}(p_{z_2}-p_{z_4})$ is omitted for symmetry reason. }
	\label{tab:cluster}
	\begin{tabular}{c|c|c|c}
		\hline\hline
		State  & $\epsilon(d)$ & $\epsilon(\mbox{effective }p)$ & Hopping  \\
		\hline
		$\{d_{x^2-y^2}, \frac{1}{2}(p_{x_1}-p_{y_2}-p_{x_3}+p_{y_4})\}$ & $\epsilon(d_{x^2-y^2})$ & $\epsilon(p_{x_1})$ + 2$V_{pp}$ & 2$V_{x^2-y^2}$ \\
		$\{d_{z^2}, \frac{1}{2}(p_{x_1}+p_{y_2}-p_{x_3}-p_{y_4})\}$ & $\epsilon(d_{z^2})$ & $\epsilon(p_{x_1})$ - 2$V_{pp}$ & 2$V_{z^2}$ \\
		$\{d_{xy}, \frac{1}{2}(p_{y_1}+p_{x_2}-p_{y_3}-p_{x_4})\}$ & $\epsilon(d_{xy})$ & $\epsilon(p_{y_1})$ - 2$V'_{pp}$ & 2$V_{xy}$ \\
		$\{d_{xz}, \frac{1}{\sqrt{2}}(p_{z_1}-p_{z_3})\}$ & $\epsilon(d_{xz})$ & $\epsilon(p_{z_1})$  & $\sqrt{2} V_{xz}$ \\
		\hline\hline
	\end{tabular}
\end{table}

\subsection*{A-3. Cluster model calculation of 3$d$ sequence}

\begin{figure*}
	\includegraphics[width=0.8\textwidth]{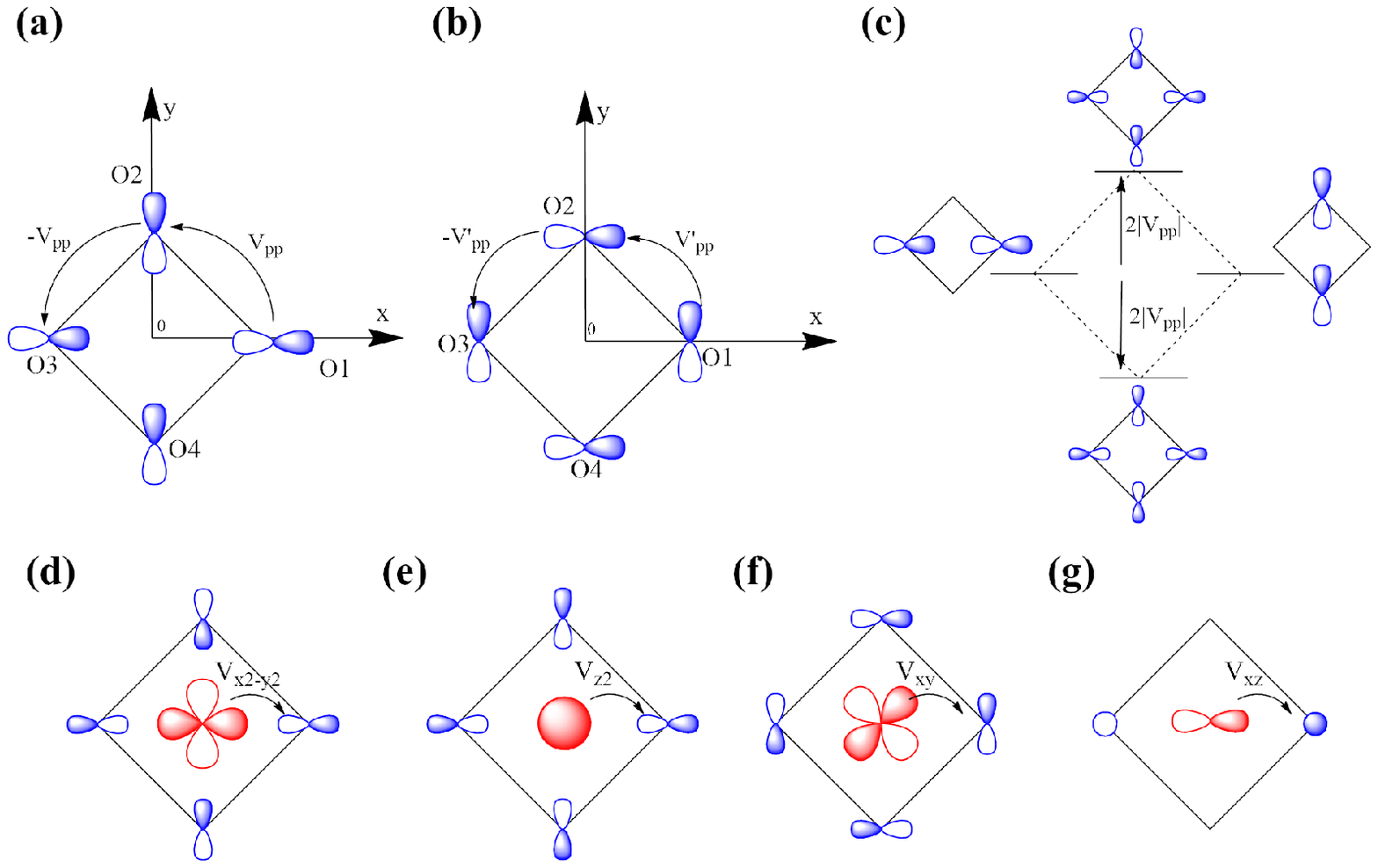}
	\caption{NiO$_4$ cluster model. (a) The linear combination of O1-$p_x$, O2-$p_y$, O3-$p_x$ and O4-$p_y$. The hopping between O1-$p_x$ and O2-$p_y$ is $V_{pp}$. (b) The linear combination of O1-$p_y$, O2-$p_x$, O3-$p_y$ and O4-$p_x$. The hopping between O1-$p_y$ and O2-$p_x$ is $V_{pp}$. (c) Energy diagram for four effective $p$ orbitals linear combined from O1-$p_x$, O2-$p_y$, O3-$p_x$ and O4-$p_y$. (d) Symmetry-allowed hopping between 3$d_{x^2-y^2}$ and $\frac{1}{2}(p_{x_1}-p_{y_2}-p_{x_3}+p_{y_4})$. The hopping between 3$d_{x^2-y^2}$ and $p_{x_1}$ is $V_{x^2-y^2}$. (e) Symmetry-allowed hopping between 3$d_{z^2}$ and $\frac{1}{2}(p_{x_1}+p_{y_2}-p_{x_3}-p_{y_4})$. The hopping between 3$d_{z^2}$ and $p_{x_1}$ is $V_{z^2}$. (f) Symmetry-allowed hopping between 3$d_{xy}$ and $\frac{1}{2}(p_{y_1}+p_{x_2}-p_{y_3}-p_{x_4})$. The hopping between 3$d_{xy}$ and $p_{y_1}$ is $V_{xy}$. (f) Symmetry-allowed hopping between 3$d_{xz}$ and $\frac{1}{\sqrt{2}}(p_{z_1}-p_{z_3})$. The hopping between 3$d_{xz}$ and $p_{z_1}$ is $V_{xz}$. Here $d_{yz}$ and $\frac{1}{\sqrt{2}}(p_{z_2}-p_{z_4}$ is omitted for symmetry reason.
	\label{fig:cluster}}
\end{figure*}

Based on the above band structure calculations and Wannier downfolding scheme, here we can calculate the effective 3d orbital sequence (which is related to the RIXS experiment) through the cluster model proposed by Eskes et al.\cite{Eskes1990}. Here we consider a NiO$_4$ cluster: four O atoms forming a square and Ni atom is at the center (see Fig. \ref{fig:cluster}(a)). We denote the bare on-site energy of 2$p_i$ as $\epsilon(p_i)$ (i=$x$, $y$, $z$) and 3$d_j$ as $\epsilon(d_j)$ (j=$z^2$, $x^2-y^2$, $xy$, $xz$, $yz$). There are three steps for this treatment.
At step-1, we start from the linear combination of $p$ on the four O atoms according to the symmetry of 3$d$ orbitals. Here we take the linear combination O1-$p_x$ (also label as ${p_{x}}_1$), O2-$p_y$, O3-$p_x$ and O4-$p_y$ as an example (Fig. \ref{fig:cluster}(a)). Suppose the hopping  between O1-$p_x$ and O2-$p_y$ is denoted by $V_{pp}$ ($V'_{pp}$ for O1-$p_y$ and O2-$p_x$ as displayed in Fig. \ref{fig:cluster}(b)). Now we consider their linear combinations, the resulting effective orbitals and onsite energies are easily calculated and the results are shown in Fig. \ref{fig:cluster}(c). The bonding orbital is expressed as $\frac{1}{2}(p_{x_1}+p_{y_2}-p_{x_3}-p_{y_4})$ with onsite energy stabilized by 2$|V_{pp}|$, so the onsite energy of this effective orbital is calculated as $\epsilon(\frac{1}{2}(p_{x_1}+p_{y_2}-p_{x_3}-p_{y_4})) = \epsilon(p_{x_1}) - 2|V_{pp}|$. The anti-bonding orbital is expressed as $\frac{1}{2}(p_{x_1}-p_{y_2}-p_{x_3}+p_{y_4})$ with onsite energy destabilized by by 2$|V_{pp}|$, so the onsite energy is $\epsilon(\frac{1}{2}(p_{x_1}-p_{y_2}-p_{x_3}+p_{y_4})) = \epsilon(p_{x_1}) + 2|V_{pp}|$. The onsite energy of left two non-bonding orbitals do not change and is $\epsilon(p_{x_1})$.

At step-2, we consider the hopping between Ni 3$d$ WFs and these effective orbitals formed by $p$. Here we take Ni 3$d_{x^2-y^2}$ for example. Suppose the hopping between O1-p$_x$ and 3$d_{x^2-y^2}$ is $V_{x^2-y^2}$ as shown in Fig. \ref{fig:cluster}(d), then the hopping between 3$d_{x^2-y^2}$ and $\frac{1}{2}(p_{x_1}-p_{y_2}-p_{x_3}+p_{y_4})$ is given by $V = 0.5*V_{x^2-y^2}*4 = 2V_{x^2-y^2}$. The other symmetry allowed hoppings are shown in Fig. \ref{fig:cluster}(e)-(g). Then, we reach the information in Tab. \ref{tab:cluster}.

At step-3, we can construct a $2 \times 2$ matrix for each 3$d$ and the corresponding effective $p$ orbitals. Diagonalizing the matrix gives two eigenvalues. Since the effective $p$ orbitals have lower onsite energies than 3$d$, the higher eigenvalue gives the onsite energy of related effective 3$d$ orbitals. For NdNiO$_2$, the parameters from above downfolding are:
$\epsilon(d_{x^2-y^2})$ = 5.57 eV, $\epsilon(d_{z^2})$ = 4.93 eV, $\epsilon(d_{xy})$ = 4.92 eV, $\epsilon(d_{xz})$ = 5.06 eV, $\epsilon(p_{x_1})$ = 1.19 eV, $\epsilon(p_{y_1})$ = 1.91 eV, $\epsilon(p_{z_1})$ = 1.96 eV, $V_{pp}$ = -0.62 eV, $V'_{pp}$ = -0.26 eV, $V_{x^2-y^2}$ = 1.28 eV, $V_{z^2}$ = -0.19 eV, $V_{xy}$ = -0.75 eV, $V_{xz}$ = -0.80 eV.
which gives effective 3$d$ sequence as:
$d_{x^2-y^2}$ (0 eV) $> d_{xy}$ (-1.53 eV) $> d_{xz}/d_{yz}$ (-1.57 eV) $> d_{z^2}$ (-2.04 eV). This is shown in Fig. 1(e) in the main text.

Importantly, the cluster method can also infer the information on the effective 3d orbitals. Here we compare NiO$_4$ cluster from NdNiO$_2$ and CuO$_4$ cluster from CaCuO$_2$ (see A-6 for more information). At shown in Tab. \ref{tab:weight}, CaCuO$_2$ is a typical charge-transfer insulator and the contribution from O 2$p$ orbitals is close to 50\%, except for $d_{z^2}$ effective orbitals. But the 2$p$ contributions are much smaller in NiO$_4$, here we take effective 3d$_{x^2-y^2}$ orbital as an example. We see the weight of O 2$p$ in this effective orbital of NdNiO$_2$ is around $23.8\%$. As a comparison, we find the weight of O 2$p$ orbital in CaCuO$_2$ is around $44.8\%$. Thus, the component of O p-orbital in NiO$_4$ is only half of that in CuO$_4$. This is one of key difference between NdNiO$_2$ and CaCuO$_2$. 
This difference is able to explain that, in the recent EELS experiment \cite{Goodge2021}, hole-doping only leads to relatively small change of O K-edge XAS spectrum in NdNiO$_2$, compared to cuprates.

\begin{table}[t]
	\caption{The weight of O $p$ orbital in each effective 3$d$ orbitals in NiO$_4$ cluster from NdNiO$_2$, compared with CuO$_4$ cluster from CaCuO$_2$. Here $d_{yz}-\frac{1}{\sqrt{2}}(p_{z_2}-p_{z_4})$ is omitted for symmetry reason. }
	\label{tab:weight}
	\begin{tabular}{c|c|c|c|c}
		\hline\hline
		O $p$ weight  & $d_{x^2-y^2}-\frac{1}{2}(p_{x_1}-p_{y_2}-p_{x_3}+p_{y_4})$ & $d_{z^2}-\frac{1}{2}(p_{x_1}+p_{y_2}-p_{x_3}-p_{y_4})$ & $d_{xy}-\frac{1}{2}(p_{y_1}+p_{x_2}-p_{y_3}-p_{x_4})$ & $d_{xz}- \frac{1}{\sqrt{2}}(p_{z_1}-p_{z_3})$ \\
		\hline
		NiO$_4$ & 23.8\% & 0.6\% & 11.9\% & 9.6\% \\
		CuO$_4$ & 44.8\% & 3.3\% & 43.7\% & 45.2\% \\
		\hline\hline
	\end{tabular}
\end{table}

\subsection*{A-5. Impurity model calculation of 3$d$ sequence}
DFT+DMFT provides an impurity model approach towards 3$d$ orbital sequence. We have performed calculations with LaNiO$_2$ and NdNiO$_2$. In both compounds, Ni-3$d$ orbitals are considered as correlated impurities. In addition, for NdNiO$_2$, two different methodologies are employed for Nd-4$f$ orbitals, namely 1) open-core treatment, and 2) correlated impurity on the equal-footing as Ni-3$d$. For each case, we have performed calculations using both $U_d=5.0$ eV, $J_d=0.8$ eV and $U_d=6.0$ eV, $J_d=0.9$ eV. For the realistic Nd calculations, $U_f=6.0$ eV, $J_f=0.7$ eV is employed for Nd-4$f$ orbitals as well. In all calculations, the continuous time quantum Monte carlo (CTQMC) impurity solver is employed. The solver samples 2$\times10^9$ steps at 116K. 

We show the crystal field splitting obtained from DFT+DMFT calculations in TAB. \ref{tab:dmftcef}. 
In all cases, the low-energy effective crystal field splitting has the same order as experimental observation.
Here we conclude the DMFT calculations give consistent results about the Ni 3$d$ sequence. 

\begin{table}[b]
	\caption{CFS obtained in DFT+DMFT calculations. In NdNiO$_2$ calculations, Nd-4f orbitals are either treated using open-core method [column NdNiO$_2$ (opencore)] or on the equal footing using CTQMC [column NdNiO$_2$ (full)]. All orbital energies are relative to d$_{x^2-y^2}$ orbitals, and all units are in eV.\label{tab:dmftcef}}
	\begin{tabular}{c|cc|cc|cc|cc|cc|cc}
		\hline\hline
		& \multicolumn{2}{c|}{LaNiO$_2$} & \multicolumn{2}{c|}{NdNiO$_2$ (opencore)} &\multicolumn{2}{c|}{NdNiO$_2$ (full)} \\
		\hline
		& \multicolumn{1}{c|}{U=5.0} & \multicolumn{1}{c|}{U=6.0} & \multicolumn{1}{c|}{U=5.0} & \multicolumn{1}{c|}{U=6.0} & \multicolumn{1}{c|}{U=5.0} & \multicolumn{1}{c|}{U=6.0} \\
		\hline
		d$_{x^2-y^2}$ & \multicolumn{1}{c|}{0.0} & \multicolumn{1}{c|}{0.0} & \multicolumn{1}{c|}{0.0} & \multicolumn{1}{c|}{0.0} & \multicolumn{1}{c|}{0.0} & \multicolumn{1}{c|}{0.0}  \\
		\hline
		d$_{xy}$ &  -1.17 &  -1.21 &  -1.28 &  -1.30 & -1.23 &  -1.25 \\
		d$_{zx/zy}$ &  -1.28 &  -1.31 & -1.33 &  -1.35 & -1.27 & -1.27 \\
		d$_{z^2}$ & -2.15 &  -2.21 &  -2.05 &  -2.08 & -1.96 & -1.96 \\
		\hline\hline
	\end{tabular}
\end{table}


\begin{figure*}
	\includegraphics[width=0.8\textwidth]{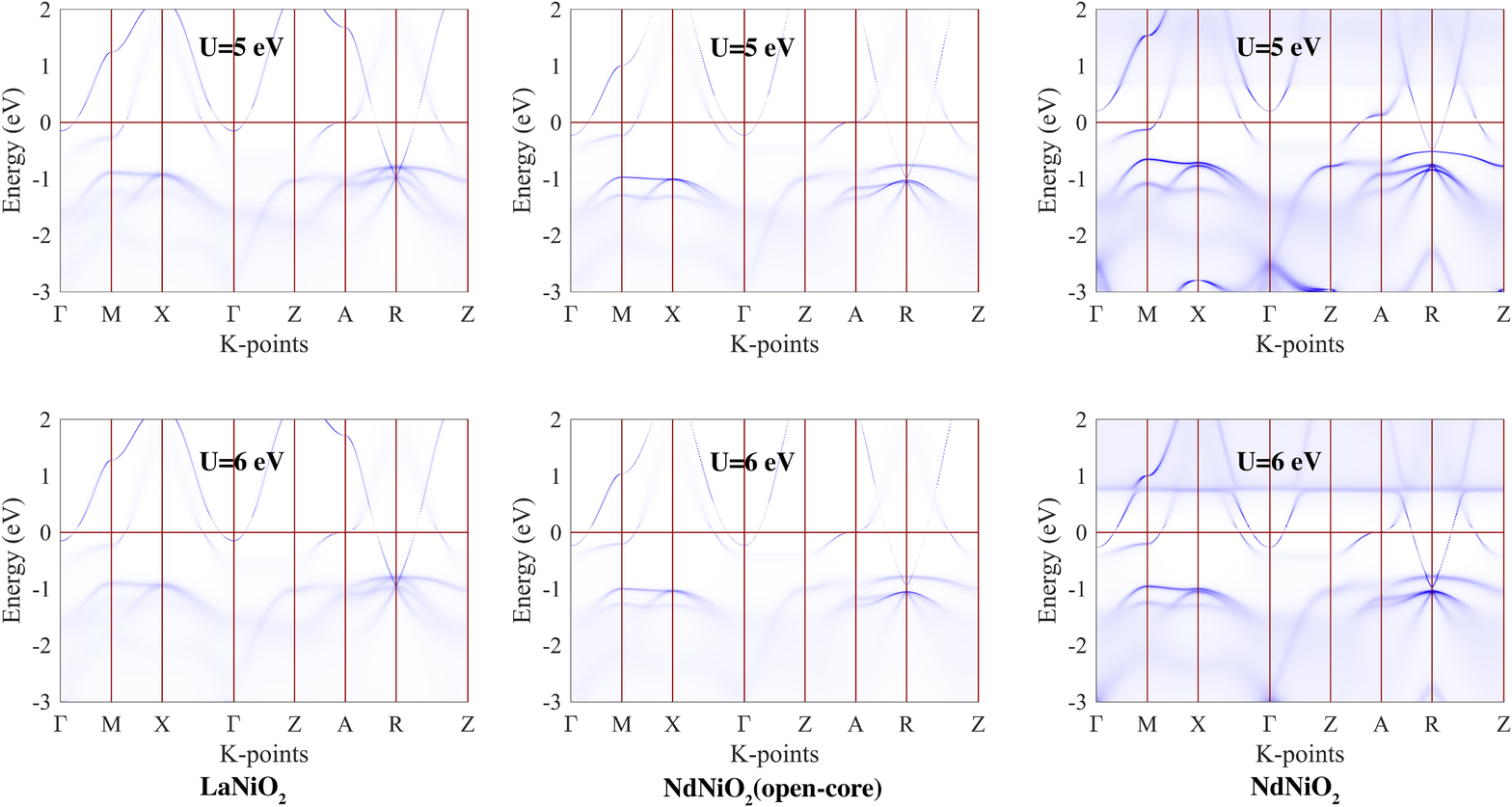}
	\caption{ Momentum-resolved spectral function from DFT+DMFT calculations for LaNiO$_2$ and NdNiO$_2$ at 116 K. 
		\label{sfig:dmft}}
\end{figure*}

\subsection*{A-6. 3$d$ sequence in CaCuO$_2$}
\begin{figure*}
	\includegraphics[width=0.8\textwidth]{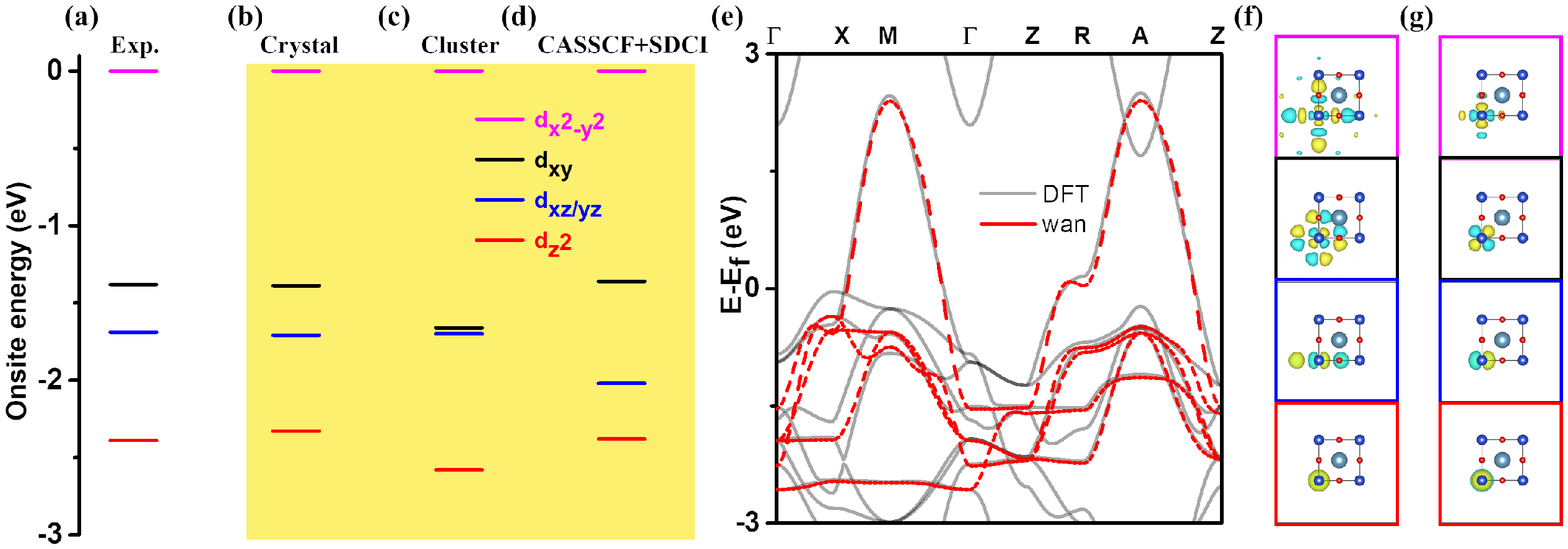}
	\caption{Onsite energy for Cu 3$d$ orbital. (a) From experimental data \cite{CASSCF2}; (b) From crystal model; (c) From cluster model; (d) From state-of-art many-body quantum chemistry method data. The onsite energy of $d_{x^2-y^2}$ is set to be zero \cite{CASSCF2}. (e) The DFT band structure and the Wannier fitted effective bands with Cu 3$d$ orbitals. Maximally localized Wannier functions for Cu 3$d$ orbital for case (f) only Cu 3$d$ orbitals and (g) both Cu 3$d$ and O 2$p$ orbitals are used in downfolding.
	\label{fig:CFS-Cu}}
\end{figure*}

To make a comparison with cuprates, here we consider the infinite layer cuprate CaCuO$_2$ \cite{Siegrist1988}. The lattice constant we use is a = b = 3.90 {\AA} and c = 3.21 {\AA}. The result is shown in Fig. \ref{fig:CFS-Cu}. Here the experimental result (Fig. \ref{fig:CFS-Cu}(a)) is taken from Hozoi et al. \cite{CASSCF2}, which the contribution of magnetic contributions are excluded \cite{cupCFS}. The CFS of the crystal model is shown in  Fig. \ref{fig:CFS-Cu}(b), which is almost the same to the experimental date in Fig. \ref{fig:CFS-Cu}(a). The Wannier fitted band structure with respect to first-principles calculation is shown in Fig. \ref{fig:CFS-Cu}(e) and the obtained WFs are shown in Fig. \ref{fig:CFS-Cu}(f), with large tails on the nearby O atoms. We can also use both Cu 3$d$ and O 2$p$ in the downfolding. Once the O 2$p$ orbitals are used, the the hybridization of O 2$p$ and Cu 3$d$ is closed and the WFs resembles atomic 3$d$ orbitals (compare Fig. \ref{fig:CFS-Cu}(g) and Fig. \ref{fig:CFS-Cu}(f)). The parameters from such downfolding in CaCuO$_2$ are:
$\epsilon(d_{x^2-y^2})$ = 2.38 eV, $\epsilon(d_{z^2})$ = 1.90 eV, $\epsilon(d_{xy})$ = 1.87 eV, $\epsilon(d_{xz})$ = 1.96 eV, $\epsilon(p_{x_1})$ = 0.59 eV, $\epsilon(p_{y_1})$ = 1.90 eV, $\epsilon(p_{z_1})$ = 1.76 eV, $V_{pp}$ = -0.64 eV, $V'_{pp}$ = -0.47 eV, $V_{x^2-y^2}$ = 1.22 eV, $V_{z^2}$ = -0.25 eV, $V_{xy}$ = -0.68 eV, $V_{xz}$ = -0.72 eV.
which allows us to calculate effective 3$d$ sequence through cluster model as:
$d_{x^2-y^2}$ (0 eV) $> d_{xy}$ (-1.66 eV) $> d_{xz}/d_{yz}$ (-1.70 eV) $> d_{z^2}$ (-2.58 eV). This is shown in Fig. \ref{fig:CFS-Cu}(c).

Hozoi et al. \cite{CASSCF2} has applied state-of-art many-body quantum chemistry methods (CASSCF+SDCI) to study the CFS and the obtained orbital order is shown in Fig. \ref{fig:CFS-Cu}(d). Therefore, all the three models give consistent CFS of Cu 3$d$ effective orbitals.

\begin{figure}
	\includegraphics[width=0.55\textwidth]{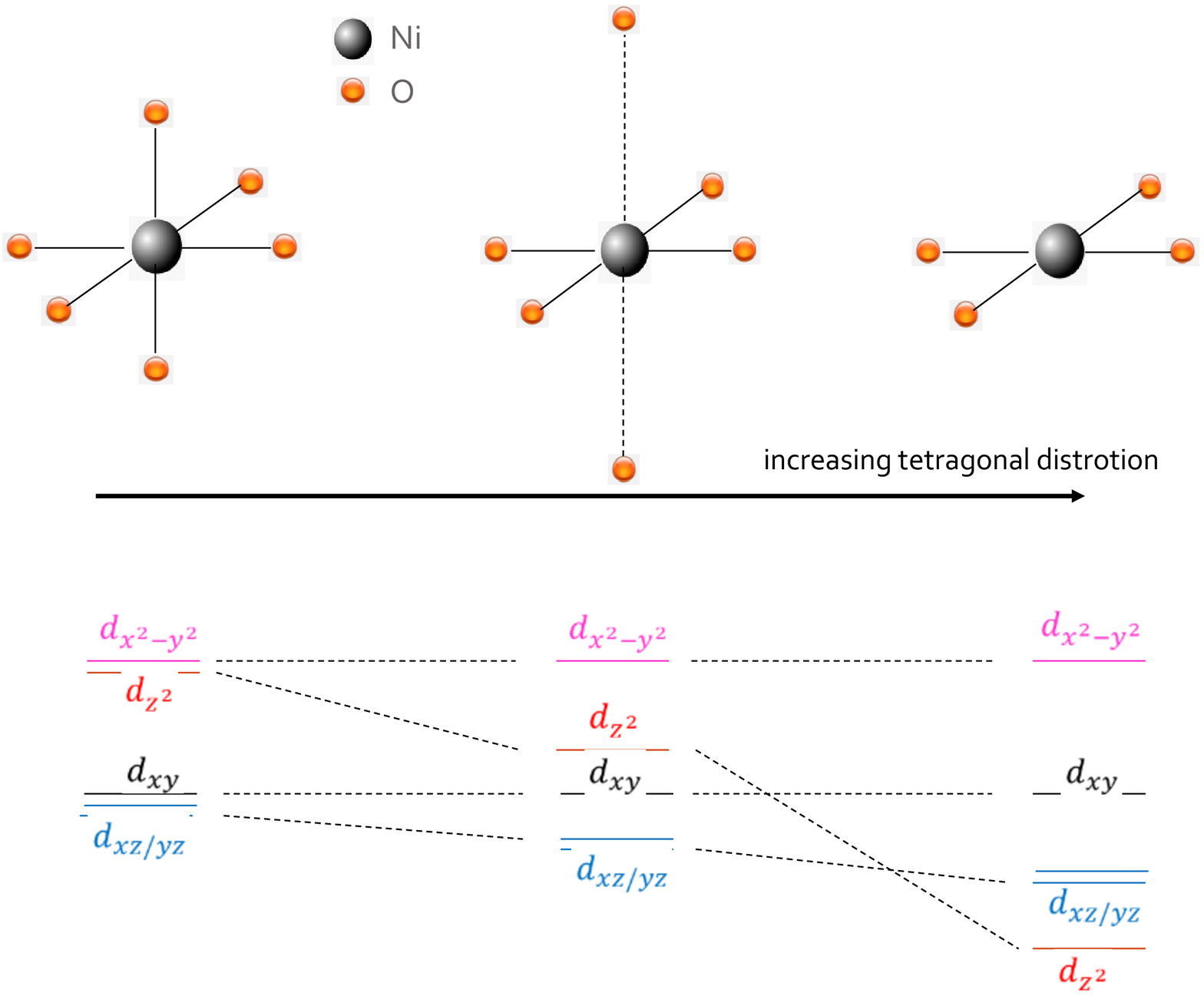}
	\caption{Schematic plot of crystal field splitting by increasing tetragonal distortion along c-axis. Different 3d orbitals are labeled by colors.   \label{sfig:cfs}}
\end{figure}

	\subsection{A-7. Physical picture for the crystal field splitting}
	In the main text, we have shown numerical results and detailed discussion on the crystal field splitting of Ni 3d orbitals. Here, we would like to provide a physical picture to understand this result.
	
	A vast of band structure calculations have shown a multi-band nature around the Fermi level \cite{bandNd0, bandNd1, bandNd2, bandNd3, multi0, multi1, multi2, multi3, multi4, multi5, multi6, multi7, multi8, multi9, multi10, cal1, cal2, cal3, cal4, cal5, cal6}. In addition to Ni 3$d_{x^2-y^2}$ band, most works \cite{multi1, multi2, multi3, multi4, multi5, multi6, multi7, multi8, multi9, multi10} prefer to use 3$d_{z^2}$ as the other target orbital based on the following two reasons: 1) 3$d_{z^2}$ contributes $\Gamma$ electron pocket; 2) in analogy to cuprates where the 3d$_{z^2} $orbital is closest to 3$d_{x^2-y^2}$.
	However, different from cuprates with $O_h$ symmetry, the point group of nickelates is reduced to $D_{4h}$, thus the crystal field splitting should be different.
	As shown in Fig. \ref{sfig:cfs}, we show a cartoon picture to understand the crystal field splitting, under the change of tetragonal distortion along c-direction.
	In contrast to $O_h$ symmetry, by removing the apical O, the out-of-plane orbitals $\{d_{z^2} , d_{xz}, d_{yz}\}$ have lower energies by extending orbital along c-axis, leaving in-plane orbitals \{$d_{x^2-y^2}$, $d_{xy}$\} relevant to the Fermi level \cite{multi0}.
	This picture applies to both infinite layer CaCuO$_2$ and NdNiO$_2$.
	This also explains why the out-of-plane 3d$_{z^2}$ is considerably lower in energy compared to the degenerate 3d$_{xy}$,3d$_{xz/yz}$ levels,
	contrary to the commonly accepted crystal field picture for a square planar coordination.

	\subsection{A-8. Parameters of two-band model}

\begin{figure*}
	\includegraphics[width=0.5\textwidth]{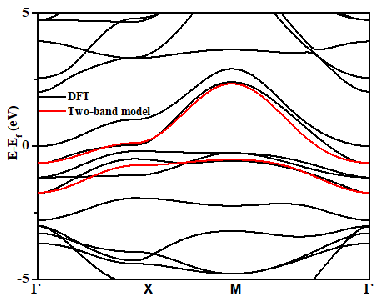}
	\caption{Band structure of two-band model and first-principles calculation for NdNiO$_2$. Here 2D Brillouin zone is used and the band structure of two-band model is shifted.  
	\label{fig:twoband}}
\end{figure*}

As Ni 3$d_{xy}$ is orthogonal to both NN and NNN 3$d_{xz}$, 3$d_{yz}$, 3$d_{z^2}$ and 3$d_{x^2-y^2}$, 3$d_{x^2-y^2}$ is orthogonal to NN and NNN 3$d_{xz}$, 3$d_{yz}$, 3$d_{xy}$, with negligible hopping to NN 3$d_{z^2}$ (0.023 eV) and orthogonal to NNN 3$d_{z^2}$, the $\{3d_{xy}, 3d_{x^2-y^2}\}$ can be regarded as orthogonal to $\{3d_{xz}, 3d_{yz}, 3d_{z^2}\}$. Such a fact allows us to separate $\{3d_{xy}, 3d_{x^2-y^2}\}$ out, which means we can directly extract the parameters related to $\{3d_{xy}, 3d_{x^2-y^2}\}$ from the crystal model. The obtaining parameters are listed in Tab. \ref{tab:occf}. With these parameters, we can recalculate the band structure as shown in Fig. \ref{fig:twoband}. The good agreement between these two indicates the validity of the model parameters.

\section*{B. Slave-boson mean-field calculation}

We use slave-boson mean-field method to deal with the interaction between electrons. The slave-boson mean-field method was first introduced to describe the un-occupied states \cite{sb01, sb02}. Then Kotliar and Ruckenstein extended the slave-boson formalism, and they used 4 slave-bosons to describe 4 different occupied states on one site \cite{Kotliar1986, Kotliar1988}. In this way, the Hubbard interaction term can be mapped to slave-boson space and simply expressed by slave-boson operators, simultaneously the hopping terms are also modified.

In this paper, we use two-orbital slave boson method to deal with interaction term \cite{Sigrist2005}. Firstly, we introduce 16 slave-bosons operators to describe 16 different occupied states in one site,
\begin{equation} \label{sb:01}
\{ e^{(\dagger)}, p_{\alpha \sigma}^{(\dagger)}, s_{\alpha}^{(\dagger)}, d_{\sigma \sigma'}^{(\dagger)}, h_{\alpha \sigma}^{(\dagger)}, f^{(\dagger)} \}
\end{equation}
where $\alpha = 1,2$ (label $d_{x^2-y^2}$ and $d_{xy}$ respectively) is band index, and $\sigma = \uparrow, \downarrow$ is spin index. These 16 slave-boson states have been listed in Tab. \ref{tab:sb1}.
\begin{table}
	\caption{The atomic states in the original model, their corresponding slave-boson states as well as the labeling of the mean fields. The site index is suppressed,$\alpha = 1,2$, and $\bar{\sigma} = \downarrow (\uparrow)$ if $\sigma = \uparrow (\downarrow)$ \cite{Sigrist2005}.}
	\begin{center}
		\begin{tabular}{|l|r|r|r|}
			\hline
			{} & Original model & Slave-boson model & Mean fields \\ \hline
			$\vert e \rangle$ & $\vert 0 \rangle$  & $e^{\dagger} \vert vac \rangle$ & $e \equiv \langle e^{(\dagger)} \rangle $  \\
			$\vert p_{\alpha \sigma} \rangle$ & $ \hat{d}_{\alpha \sigma} ^{\dagger} \vert 0 \rangle$ &  $ p_{\alpha \sigma} ^{\dagger} \hat{d}_{\alpha \sigma}^{\dagger} \vert vac \rangle$ & $p_{\alpha \sigma} \equiv \langle p_{\alpha \sigma}^{(\dagger)} \rangle$  \\
			$\vert s_{\alpha} \rangle$ & $ \hat{d}_{\alpha \uparrow} ^{\dagger} \hat{d}_{\alpha \downarrow} ^{\dagger} \vert 0 \rangle$ &  $ s_{\alpha} ^{\dagger} \hat{d}_{\alpha \uparrow}^{\dagger} \hat{d}_{\alpha \downarrow}^{\dagger} \vert vac \rangle$ & $s_{\alpha} \equiv \langle s_{\alpha}^{(\dagger)} \rangle$  \\
			$\vert d_{\sigma \sigma} \rangle$ & $ \hat{d}_{1 \sigma} ^{\dagger} \hat{d}_{2 \sigma} ^{\dagger} \vert 0 \rangle$ &  $ d_{\sigma \sigma} ^{\dagger} \hat{d}_{1 \sigma}^{\dagger} \hat{d}_{2 \sigma}^{\dagger} \vert vac \rangle$ & $d_{\sigma \sigma} \equiv \langle d_{\sigma \sigma}^{(\dagger)} \rangle$  \\
			$\vert d_{\sigma \bar{\sigma}} \rangle$ & $ \hat{d}_{1 \sigma} ^{\dagger} \hat{d}_{2 \bar{\sigma}} ^{\dagger} \vert 0 \rangle$ &  $ d_{\sigma \bar{\sigma}} ^{\dagger} \hat{d}_{1 \sigma}^{\dagger} \hat{d}_{2 \bar{\sigma}}^{\dagger} \vert vac \rangle$ & $d_{\sigma \bar{\sigma}} \equiv \langle d_{\sigma \bar{\sigma}}^{(\dagger)} \rangle$  \\
			$\vert h_{1 \sigma} \rangle$ & $ \hat{d}_{1 \sigma} ^{\dagger} \hat{d}_{2 \uparrow} ^{\dagger} \hat{d}_{2 \downarrow} ^{\dagger}\vert 0 \rangle$ &  $ h_{1 \sigma} ^{\dagger} \hat{d}_{1 \sigma}^{\dagger} \hat{d}_{2 \uparrow}^{\dagger} \hat{d}_{2 \downarrow}^{\dagger} \vert vac \rangle$ & $h_{1 \sigma} \equiv \langle h_{1 \sigma}^{(\dagger)} \rangle$  \\
			$\vert h_{2 \sigma} \rangle$ & $ \hat{d}_{1 \uparrow} ^{\dagger} \hat{d}_{1 \downarrow} ^{\dagger} \hat{d}_{2 \sigma} ^{\dagger} \vert 0 \rangle$ &  $ h_{2 \sigma} ^{\dagger} \hat{d}_{1 \uparrow}^{\dagger} \hat{d}_{1 \downarrow}^{\dagger} \hat{d}_{2 \sigma}^{\dagger} \vert vac \rangle$ & $h_{2 \sigma} \equiv \langle h_{2 \sigma}^{(\dagger)} \rangle$  \\
			$\vert f \rangle$ & $ \hat{d}_{1 \uparrow} ^{\dagger} \hat{d}_{1 \downarrow} ^{\dagger} \hat{d}_{2 \uparrow} ^{\dagger} \hat{d}_{2 \downarrow} ^{\dagger} \vert 0 \rangle$  & $f^{\dagger} \hat{d}_{1 \uparrow}^{\dagger} \hat{d}_{1 \downarrow}^{\dagger} \hat{d}_{2 \uparrow}^{\dagger} \hat{d}_{2 \downarrow}^{\dagger} \vert vac \rangle$ & $f \equiv \langle f^{(\dagger)} \rangle $  \\
			\hline
		\end{tabular}
	\end{center}
	\label{tab:sb1}
\end{table}
The introduction of 16 slave-bosons enlarge the Hilbert space to an unphysical one, so we need some local constraints to form a physical space. Summing up all slave boson operators we define
\begin{equation}
\hat{I}_i = e_i^{\dagger} e_i + \sum_{\substack{\alpha \sigma}} (p_{i \alpha \sigma}^{\dagger} p_{i \alpha \sigma} + h_{i \alpha \sigma}^{\dagger} h_{i \alpha \sigma}) + \sum_{\substack{\alpha}} s_{i \alpha}^{\dagger} s_{i \alpha} + \sum_{\substack{\sigma \sigma'}} d_{i \sigma \sigma'}^{\dagger} d_{i \sigma \sigma'} + f_i^{\dagger} f_i
\end{equation}
And define the operators
\begin{equation}
\hat{Q}_{i 1 \sigma} = p_{i 1 \sigma}^{\dagger} p_{i 1 \sigma} +s_{i 1}^{\dagger} s_{i 1} +\sum_{\substack{\sigma'}} d_{i \sigma \sigma'}^{\dagger} d_{i \sigma \sigma'} + h_{i 1 \sigma}^{\dagger} h_{i 1 \sigma} +  \sum_{\substack{\sigma}} h_{i 2 \sigma}^{\dagger}
h_{i 2 \sigma} + f_i^{\dagger} f_i
\end{equation}
\begin{equation}
\hat{Q}_{i 2 \sigma} = p_{i 2 \sigma}^{\dagger} p_{i 2 \sigma} +s_{i 2}^{\dagger} s_{i 2} +\sum_{\substack{\sigma'}} d_{i \sigma' \sigma}^{\dagger} d_{i \sigma' \sigma} + h_{i 2 \sigma}^{\dagger} h_{i 2 \sigma} +  \sum_{\substack{\sigma}} h_{i 1 \sigma}^{\dagger}
h_{i 1 \sigma} + f_i^{\dagger} f_i
\end{equation}
Thus, the physical subspace is given by two kinds local constraints:
\begin{equation} \label{sb02}
\hat{I}_i - 1 \equiv 0
\end{equation}
\begin{equation} \label{sb03}
\hat{f}_{i \alpha \sigma}^{\dagger} \hat{f}_{i \alpha \sigma} - \hat{Q}_{i \alpha \sigma} \equiv 0
\end{equation}
These constraints ensure that the slave-boson states form a complete set in the physical local Hilbert space of the slave-boson model. The first relation (\ref{sb02}) represents the completeness of the boson operators, i.e., the total probability of slave-bosons on one site is 1.
The second relation (\ref{sb03}) is similar to the conservation of the number of particles, i.e., the charge of bosons should be equal to the electron number. Therefore, we have to ensure that in the physical subspace the operators $\hat{Q}_{i \alpha \sigma}$ are identical to the operators $\hat{f}_{i \alpha \sigma}^{\dagger} \hat{f}_{i \alpha \sigma}$. Using these constraints and neglecting the spin-flip and pair-hopping term in Hund coupling, the interaction term becomes
quadratic in the boson operators :
\begin{equation} \label{009}
\begin{aligned}
\hat{H}_{int} = &\sum_{\substack{i}} \Bigg \{ U\sum_{\substack{\alpha}} s_{i \alpha}^{\dagger} s_{i \alpha} + (U + 2U' -J_H)\sum_{\substack{\alpha \sigma}}  h_{i \alpha \sigma}^{\dagger} h_{i \alpha \sigma} \\
&+ (U'-J_H) \sum_{\substack{\sigma}} d_{i \sigma \sigma}^{\dagger} d_{i \sigma \sigma} + U' \sum_{\substack{\sigma}} d_{i \sigma \bar{\sigma}}^{\dagger} d_{i \sigma \bar{\sigma}} \\
& +2(U+2U'-J_H) f_i^{\dagger} f_i \Bigg \}
\end{aligned}
\end{equation}
But the hopping term become more complex by the correction of slave-bosons. There is a mapping for hopping term:
\begin{equation}
\begin{aligned}
\hat{d}_{i \alpha \sigma} & \to \tilde{z}_{i \alpha \sigma} \hat{d}_{i \alpha \sigma} \\
\hat{d}_{i \alpha \sigma}^{\dagger} & \to \hat{d}_{i \alpha \sigma}^{\dagger} \tilde{z}_{i \alpha \sigma}^{\dagger}
\end{aligned}
\end{equation}
where
\begin{equation} \label{011}
\begin{aligned}
\tilde{z}_{i \alpha \sigma} =& (1-\hat{Q}_{i \alpha \sigma})^{-1/2} z_{i \alpha \sigma} \hat{Q}_{i \alpha \sigma}^{-1/2} \\
z_{i 1 \sigma} =& e_i^{\dagger} p_{i 1 \sigma}
+ p_{i 1 \bar{\sigma}}^{\dagger} s_{i 1}
+ p_{i 2 \sigma}^{\dagger} d_{i \sigma \sigma}
+ p_{i 2 \bar{\sigma}}^{\dagger} d_{i \sigma \bar{\sigma}} \\
&+ s_{i 2}^{\dagger} h_{i 1 \sigma}
+ d_{i \bar{\sigma} \sigma}^{\dagger} h_{i 2 \sigma}
+ d_{i \bar{\sigma} \bar{\sigma}}^{\dagger} h_{i 2 \bar{\sigma}}
+ h_{i 1 \bar{\sigma}}^{\dagger} f_{i} \\
z_{i 2 \sigma} =& e_i^{\dagger} p_{i 2 \sigma}
+ p_{i 2 \bar{\sigma}}^{\dagger} s_{i 2}
+ p_{i 1 \sigma}^{\dagger} d_{i \sigma \sigma}
+ p_{i 1 \bar{\sigma}}^{\dagger} d_{i \bar{\sigma} \sigma } \\
&+ s_{i 1}^{\dagger} h_{i 2 \sigma}
+ d_{i \sigma \bar{\sigma}}^{\dagger} h_{i 1 \sigma}
+ d_{i \bar{\sigma} \bar{\sigma}}^{\dagger} h_{i 1 \bar{\sigma}}
+ h_{i 2 \bar{\sigma}}^{\dagger} f_{i} \\
\end{aligned}
\end{equation}
The “z-operators” keep track of the bosons during hopping processes and the choice of the “z-operators” is not unique.
In our choice, the hopping term canbe written as :
\begin{equation}
\hat{H}_{TB} = \sum_{i,\alpha,\sigma} \epsilon_\alpha \hat{n}_{i\alpha\sigma} + \sum_{<i,j>,\alpha,\sigma}(t_\alpha   \hat{f}^{\dagger}_{i\alpha\sigma} \tilde{z}_{i \alpha \sigma}^\dagger \tilde{z}_{j \alpha \sigma}  \hat{f}_{j\alpha\sigma} + h.c.)
\end{equation}
Only the hopping term is corrected here, and the on-site energy is unchanged. The saddle-point approximation is equivalent to a mean-field approximation where the Bose fields and Lagrange multipliers are treated as static and homogeneous fields \cite{Kotliar1986, Kotliar1988, Sigrist2005}. Thus, this approximation consists essentially in replacing the creation and annihilation operators of the slave bosons by site independent c-numbers which can be chosen to be real. So the interaction term is only depended on these c-number:
\begin{equation}
\begin{aligned}
H_{int} =& NU(s_{1}^2 + s_{2}^2) + N(U + 2U' -J_H)(h_{1 \uparrow}^2 + h_{1 \downarrow}^2 +h_{2 \uparrow}^2 +h_{2 \downarrow}^2)\\
&+ N(U'-J_H) (d_{\uparrow \uparrow}^2 + d_{\downarrow \downarrow}^2) + NU' (d_{\uparrow \downarrow}^2 + d_{\downarrow \uparrow}^2) \\
&+ 2N(U+2U'-J_H) f^2
\end{aligned}
\end{equation}
And we define a new factor $q_{\alpha \sigma} = < \tilde{z}_{\alpha \sigma}^{\dagger} \tilde{z}_{\alpha \sigma} > $ to describe the corraction of hopping. In our choice, q-factor is defined as a real number from 0 to 1. And as Hubbard interaction strength increase, q-factors decrease. If the band is half-filling, q-factor will become zero which means a Mott insulator. And the hopping term can be written as
\begin{equation}
\hat{H}_{TB} = \sum_{i,,\sigma} \epsilon_\alpha \hat{n}_{i\alpha\sigma} + \sum_{<i,j>,\alpha,\sigma}(t_\alpha q_{\alpha \sigma}  \hat{d}^{\dagger}_{i\alpha\sigma} \hat{d}_{j\alpha\sigma} + h.c.)
\end{equation}
It can be easily seen that the role of the slave-bosens is to renormalize the electronic hopping strength. These c-numbers of slave-bosons canbe solved by minimization of free energy.
Next, we need to simplify our model, because the 16 slave-bosen parameters are not easy to solve, even in mean-field level.
Let's analyze the real situation of our system. In our system, when we consider Ni 3$d_{x^2-y^2}$ and 3$d_{xy}$ orbitals, the electron number of density on every site is between 2 and 3. Moreover, there is a large crystal field split between two orbitals. And the Hubbard interaction strength $U$ in two orbits is very large, which prevent two electrons occupy the same orbit on the same site. In addition, we assume that system is spin-degenerate. So we believe that the probability of some electron occupied states is very small, such as empty, single and four occupied states, and these states canbe ignored in our case, so it is only left four effective occupied states in our model, they are $\{s,d_1,d_2,h_1\}$. $s$ means there are two electrons in lower band ($s = s_2$). $d_1$ means the each band have one electron ($d_1 = d_{\uparrow \uparrow} = d_{\downarrow \downarrow}$), and the two electrons are arranged paramagnetically. Relatively, $d_2$ means two electrons are antiferromagnetic arranged ($d_2 = d_{\uparrow \downarrow} = d_{\downarrow \uparrow}$). $h_1$ means there are two electrons in lower band and one electron in upper band ($h_1 = h_{1 \uparrow} = h_{1 \downarrow}$). So under this condition, the interaction term is simplified to
\begin{equation} \label{eq:int.01}
\begin{aligned}
H_{int} =& NUs^2 + 2N(U'-J_H)d_1^2 + 2NU'd_2^2 + 2N(U+2U'-J_H)h_1^2
\end{aligned}
\end{equation}
and the completeness relationship of the slave bosons Eq. (\ref{sb02}) and the conservation of the number of bosons and electrons Eq. (\ref{sb03}) can be expressed as
\begin{equation} \label{eq:completeness relationship}
\begin{aligned}
1 &= s^2 + 2(d_1^2 + d_2^2 + h_1^2)
\end{aligned}
\end{equation}
\begin{equation} \label{eq:n1}
\begin{aligned}
n_1 &= 2(d_1^2 + d_2^2 + h_1^2)
\end{aligned}
\end{equation}
\begin{equation} \label{eq:n2}
\begin{aligned}
n_2 &= 2(s^2 + d_1^2 + d_2^2 + 2h_1^2)
\end{aligned}
\end{equation}
With these constraints, q-factor of hopping term also simplied:
\begin{equation}
\begin{aligned}
q_1 &=
\frac{2(1-2d_1^2-2d_2^2-s_1^2) s^2}{(1-s^2)(1+s^2)}\\
q_2 &= \frac{(1-2d_1^2-2d_2^2-s^2) (d_1+d_2)^2}{2(d_1^2+d_2^2)(1-d_1^2-d_2^2)}
\end{aligned}
\end{equation}
For the modification of superconductivity, we discuss in next section.

\section{C. Bogoliubov-de Gennes equation and superconductivity}
We use the Bogoliubov-de Gennes (BdG) method to deal with the superconductivity. Firstly, It should be noticed that there is no coupling between two bands after slave-boson mean-field approximation. So we can treat the two energy bands respectively as single band. Thus, the single band Hamiltonian canbe written as:
\begin{equation} \label{SC01}
\begin{aligned}
\hat{H}_\alpha &= -\sum_{i,\sigma} \mu_\alpha \hat{n}_{i \alpha \sigma} + t_\alpha q_\alpha \sum_{<i,j>,\sigma}(  \hat{d}^{\dagger}_{i \alpha \sigma} \hat{d}_{j \alpha \sigma} + h.c.) + \frac{1}{4} J_\alpha \sum_{\langle ij \rangle} (4 \mathbf{S}_{i \alpha} \cdot \mathbf{S}_{j \alpha} - n_{i \alpha} n_{j \alpha})
\end{aligned}
\end{equation}
where $\mu_\alpha = \mu - \epsilon_\alpha$ is the chemical potential of band $\alpha$. And we have ignored the Hund coupling, because the Hund coupling term only depends on slave-boson mean-field parameters. For simplicity, we absorb the coefficient $\frac{1}{4}$ into $J_\alpha$ in the following text, that is $J_\alpha = \frac{1}{4} J_\alpha$  \\
Next, by using mean-field approximation and translating it into k-space, we get new Hamiltonian in mean-field level \cite{Kotliar1988} :
\begin{equation} \label{SC02}
\begin{aligned}
\hat{H}_\alpha &=  \sum_{\substack{k \sigma}}  [-2 (K+t_\alpha) (\cos k_x + \cos k_y) -\mu_\alpha] \hat{d}_{k \alpha \sigma}^\dagger \hat{d}_{k \alpha \sigma}  \\
& - \sum_{\substack{k}} (
\Delta_d^* \eta_k \hat{d}_{-k \alpha \downarrow} \hat{d}_{k \alpha \uparrow}
+ \Delta_d \eta_k  \hat{d}_{k \alpha \uparrow}^\dagger \hat{d}_{-k \alpha \downarrow}^\dagger ) \\
&+ \frac{N |\Delta_d|^2}{3J_\alpha} + \frac{4 N K^2}{3J_\alpha} + 2J_\alpha N n_\alpha (1-2n_\alpha)  \\
\end{aligned}
\end{equation}
where $\eta_k = \cos k_x - \cos k_y$ , asuming a d-wave symmetry pairing, and we define order parameters as
\begin{equation}
\begin{aligned}
\Delta_d = & \frac{3J_\alpha}{N}  \sum_{\substack{k}} \eta_k \langle \hat{d}_{-k \alpha \downarrow} \hat{d}_{k \alpha \uparrow} \rangle \\
K = & \frac{3J_\alpha}{2N} \sum_{\substack{k}} (\cos k_x + \cos k_y) \langle \hat{d}_{k \alpha \sigma}^\dagger \hat{d}_{k \alpha \sigma} \rangle
\end{aligned}
\end{equation}

Before to solve the Hamiltonian (\ref{SC02}), we need to consider the influence by slave-bosons. As mentioned in the previous section, slave boson method modifies hopping term with q-factors, so we also introduce the q-factors into superconductivity:
\begin{equation} \label{SC03}
\begin{aligned}
\hat{H}_\alpha &=  \sum_{\substack{k \sigma}} [-2 q_\alpha (K + t_\alpha) (\cos k_x + \cos k_y) - \mu_\alpha ] \hat{d}_{k \alpha \sigma}^\dagger \hat{d}_{k \alpha \sigma} \\
&- q_\alpha \sum_{\substack{k}} (
\Delta_d^* \eta_k \hat{d}_{-k \alpha \downarrow} \hat{d}_{k \alpha \uparrow}
+ \Delta_d \eta_k  \hat{d}_{k \alpha \uparrow}^\dagger \hat{d}_{-k \alpha \downarrow}^\dagger  ) \\
&+ \frac{N |\Delta_d|^2}{3J_\alpha} + \frac{4 N K^2}{3J_\alpha} + 2J_\alpha Nn_\alpha (1-2n_\alpha)  \\
\end{aligned}
\end{equation}
And order parameters also be modified by q-factors:
\begin{equation}
\begin{aligned}
\Delta_d = & \frac{3J_\alpha q_\alpha}{N}  \sum_{\substack{k}} \eta_k \langle \hat{d}_{-k \alpha \downarrow} \hat{d}_{k \alpha \uparrow} \rangle \\
K = & \frac{3J_\alpha q_\alpha}{2N} \sum_{\substack{k}} (\cos k_x + \cos k_y) \langle \hat{d}_{k \alpha \sigma}^\dagger \hat{d}_{k \alpha \sigma} \rangle
\end{aligned}
\end{equation}
So we can conclude that if the electron kinetic energy is zero, there is no superconductivity in system.
To slove the Hamiltonian (\ref{SC03}), we introduce the BdG method. The Bogoliubov transformation of Fermion operator is
\begin{equation} \label{BdG01}
\begin{cases}
\hat{d}_{k \alpha \uparrow}^\dagger = \sum\limits_n^{'}(u_{nk}^{*} \hat{\gamma}_{nk,\uparrow}^\dagger + v_{nk} \hat{\gamma}_{n,-k,\downarrow})\\
\hat{d}_{-k \alpha \downarrow} = \sum\limits_n^{'}(u_{nk} \hat{\gamma}_{n,-k,\downarrow} - v_{nk}^{*} \hat{\gamma}_{nk,\uparrow}^\dagger)
\end{cases}
\end{equation}
where the $\prime$ over the summation means only sum with positive energy eigenvalue, $\hat{\gamma}_{nk \sigma}^\dagger$ and $\hat{\gamma}_{nk \sigma}$ are the quasi-particle generation and annihilation operators and they satisfy the anticommutation relation. By using Bogoliubov transformation, the diagonalized  Hamiltonian canbe written as
\begin{equation} \label{BdG02}
\hat{H}_{eff} = E_g + \sum\limits_{n,k,\sigma}^{'}  \hat{\gamma}_{nk \sigma}^\dagger \hat{\gamma}_{nk \sigma}
\end{equation}
We mark a new kinetic energy parameter as $\varepsilon_k = -2(K+t_\alpha)(\cos k_x + \cos k_y)$ for simplicity. And the commutation relation between the creation (annihilation) operator of electrons and the system Hamiltonian is
\begin{equation} \label{BdG03}
\begin{aligned}
\lbrack \hat{d}_{k \alpha \uparrow}^\dagger , \hat{H}_\alpha \rbrack &= -(\varepsilon_k -\mu_\alpha) \hat{d}_{k \alpha \uparrow}^\dagger + \Delta_d^* \eta_k \hat{d}_{-k \alpha \downarrow} \\
\lbrack \hat{d}_{-k \alpha \downarrow} , \hat{H}_\alpha \rbrack &= (\varepsilon_k -\mu_\alpha)
\hat{d}_{-k \alpha \downarrow} +  \Delta_d^* \eta_k \hat{d}_{k \alpha \uparrow}^\dagger
\end{aligned}
\end{equation}
Substitute Eq. (\ref{BdG01}) and Eq. (\ref{BdG02}) into Eq. (\ref{BdG03}) and compare the coefficients of the quasi-particle operators on both sides of the equation. We can get the coupled equations of coefficients $\{ u_{nk}, v_{nk} \}$:
\begin{equation} \label{BdG04}
E_{nk}
\begin{pmatrix}
u_{nk} \\
v_{nk}  \\
\end{pmatrix}
=
\begin{pmatrix}
\tilde{\varepsilon}_k - \mu_\alpha   &   \Delta_d \eta_k  \\
\Delta_d^*\eta_k  & -\tilde{\varepsilon}_k + \mu_\alpha   \\
\end{pmatrix}
\begin{pmatrix}
u_{nk} \\
v_{nk} \\
\end{pmatrix}
\end{equation}
And the self-consistent equations of mean-field order parameter and number of density can be written as
\begin{equation} \label{BdG05}
\begin{aligned}
\Delta_d =&  - \frac{3J_\alpha q_\alpha}{N}  \sum_{\substack{n k}} \eta_k u_{nk} v_{nk}^*n_F(E_{nk}) \\
K = & \frac{3J_\alpha q_\alpha}{2N} \sum_{\substack{nk}} (\cos k_x + \cos k_y) |u_{nk}|^2 n_F(E_{nk}) \\
n_\alpha =& \frac{2}{N} \sum_{\substack{nk}} |u_{nk}|^2 n_F(E_{nk})
\end{aligned}
\end{equation}
where $n_F(E_{nk})$ is the Fermi-Dirac distribution with energy $E_{nk}$. Self-consistent iteration Eq. (\ref{BdG04}) and Eq. (\ref{BdG05}), we can get the mean-field order parameters. The last point to mention is that although the superconducting order parameter $\Delta_d$ is written here as a complex number, it is actually a real number under the conditions we consider.

\section{D. Discussion on R 5$d$ electrons}

In the main text, we only keep two correlated Ni 3$d$ orbitals in the construction of effective model, by neglecting 5$d$ electron band from rare-earth element. Here we present several remarks, and explain why we discard R 5$d$ electron band in the effective model:
\begin{enumerate}
	\item We notice that, in a recent experiment on Nd$_6$Ni$_5$O$_{8}$ compound \cite{Pan2021,Botana2021} (which hosts a 3d$^{8.8}$ configuration, named n=5 in series $R_{n+1}Ni_nO_{2n+1}$), superconductivity survives and shows very similar behavior with infinite nickelates ($n=\infty$). However, the 5d band around the Fermi level of Nd$_6$Ni$_5$O$_{8}$ compound is totally different from that in infinite nickelates: Instead of a 5$d_{z^2}$ band around $\Gamma$ point, Nd$_6$Ni$_5$O$_{8}$ shows a 5$d_{xy}$ band around M point. This dramatic 5d band difference leads to the similar superconducting behavior strongly supports that 5d band from rare-earth is irrelevant to superconductivity.
	\item In infinite-layer nickelates, the band around the $\Gamma$ point is mainly made of 5$d_{z^2}$ orbital, which has sizable hybridization with Ni 3$d_{z^2}$ orbital. However, as we elucidated in the main text, both RIXS experiment \cite{CFS0} and our calculations show  Ni 3$d_{z^2}$ orbital is deeply below the Fermi level and hardly contributes to the physics in the NiO$_2$ plane. Thus, if we focus on the nature of superconductivity, that is believed to occur in the  NiO$_2$ plane, it is reasonable to neglect 5$d$ band around $\Gamma$ point in the effective model.
	\item Under hole-doping, 5$d$ band around the $\Gamma$ point quickly vanishes (or its contribution around the Fermi level vanishes), which implies this 5$d$ band is irrelevant to the superconducting nature \cite{Liu2021}.
\end{enumerate}

Based on the above reasons, we speculate that the 5$d$ band from rare-earth element contributes to modify the electron correlations on Ni 3$d_{z^2}$ orbital through the hybridization effect, and to serve as a charge reservoir.
In this regarding, the existence of a R-5$d$ band can explain that the charge carriers changes from electron-like to hole-like upon hole doping in the Hall measurement. That is,
The existence of R-5$d$ band contributes electron-like carriers in the parent compound, and then these electron-like carriers continues reduce upon hole doping. At the critical doping level, contribution of R-5$d$ band around Fermi level vanishes, so that the carrier type becomes hole-like. This is confirmed in many DFT calculations, e.g. \cite{Liu2021}

\end{document}